\documentclass[12pt]{article}
\usepackage{amssymb,amsmath,graphicx,mathtools}
\makeatletter
\usepackage{empheq}
\usepackage{jheppub}
\usepackage{hyperref}
\usepackage[none]{hyphenat}
\usepackage[utf8]{inputenc}
\usepackage{tikz} 
\usetikzlibrary{hobby} 
\usetikzlibrary{decorations.markings}
 \usetikzlibrary{decorations.pathreplacing}

\newcommand\q{0.4} % graph cell size

\definecolor{b_blue}{rgb}{0.01, 0.28, 1.0}

\newcommand{\puncture}[2]{
    \draw[draw=black, fill=red] (#1*\q, #2*\q) circle (1.5pt); %1
}

\newcommand{\handle}[2]{    
    \draw[draw=white, fill=white] (#1*\q, #2*\q)--(#1*\q, #2*\q + 1.5*\q) to[out=60,in=120] (#1*\q + 1.5*\q, #2*\q + 1.5*\q)--(#1*\q + 1.5*\q, #2*\q)--(#1*\q + \q, #2*\q)--(#1*\q + \q, #2*\q + 1.3*\q) to[out=140,in=60] (#1*\q + 0.5*\q, #2*\q + 1.3*\q)--(#1*\q + 0.5*\q, #2*\q)--(#1*\q, #2*\q);
    
    \draw[draw=blue] (#1*\q, #2*\q)--(#1*\q, #2*\q + 1.5*\q) to[out=60,in=120] (#1*\q + 1.5*\q, #2*\q + 1.5*\q)--(#1*\q + 1.5*\q, #2*\q);
    \draw[draw=blue] (#1*\q + \q, #2*\q)--(#1*\q + \q, #2*\q + 1.3*\q) to[out=140,in=60] (#1*\q + 0.5*\q, #2*\q + 1.3*\q)--(#1*\q + 0.5*\q, #2*\q);
}

\@addtoreset{equation}{section}
\def\section{\@startsection {section}{1}{\z@}{-2.5ex plus -1ex minus
 -.2ex}{1.3ex plus .2ex}{\large\bf}}
\def\subsection{\@startsection{subsection}{2}{\z@}{-2.25ex plus%
 -1ex minus -.2ex}{0.5ex plus .2ex}{\bf}}

\advance \voffset by -0.8in
\advance \hoffset by -0.6in
\def\calC{{\mathcal C}}
\def\calA{{\mathcal A}}
\def\calB{{\mathcal B}}
\def\calF{{\mathcal F}}
\def\calG{{\mathcal G}}
\def\calJ{{\mathcal J}}
\def\calP{{\mathcal P}}
\def\calR{{\mathcal R}}
\newcommand{\PT}{S}
\newcommand{\pt}{s}
\newcommand{\Rsix}{\mathbb{R}^6}

\def\P{Q}

\def\sign{{+}}
\def\osign{{-}}

\def\ba{{\mbox{\boldmath $a$}}}
\def\bx{{\mbox{\boldmath $x$}}}

\def\bj{{\mbox{\boldmath $j$}}}

\def\bs{{\mbox{\boldmath $s$}}}
\def\bp{{\mbox{\boldmath $p$}}}

\def\bq{{\mbox{\boldmath $q$}}}

\def\bn{{\mbox{\boldmath $n$}}}

\def\bv{{\mbox{\boldmath $v$}}}
\def\bu{{\mbox{\boldmath $u$}}}

\def\bpm{\begin{pmatrix}}
\def\epm{\end{pmatrix}}
\newcommand{\ca}{\mathfrak{a}} 
\newcommand{\cb}{\mathfrak{b}}

\newcommand{\cg}{\mathfrak{g}}
\newcommand{\ch}{\mathfrak{h}}
\newcommand{\ci}{\mathfrak{i}} 
\newcommand{\cj}{\mathfrak{j}} 
\newcommand{\ck}{\mathfrak{k}}

\newcommand{\RR}{\mathbb{R}}
\newcommand{\CC}{\mathbb{C}}

\newcommand{\tens}{\mathop{\otimes}}
\newcommand{\la}{{\triangleright}}

\newcommand{\ad}{{\rm ad}}
\newcommand{\Ad}{{\rm Ad}}

\def\beq{\begin{equation}}
\def\eeq{\end{equation}}
\def\bee{\begin{equation}}
\def\eee{\end{equation}}
\def\bea{\begin{eqnarray}}
\def\eea{\end{eqnarray}}

\newtheorem{theorem}{Theorem}[section]

\newtheorem{definition}[theorem]{Definition}

\def\rcross{{\triangleright\!\!\!<}}

\def\dcross{{\bowtie}}

\def\rbiprod{{\cdot\kern-.33em\triangleright\!\!\!<}}
\def\lbiprod{{>\!\!\!\triangleleft\kern-.33em\cdot}}

\makeatletter
\gdef\@fpheader{}
\makeatother

\begin{document}

\parskip 5pt
\parindent 8pt

\begin{center}

{\Large \bf  A Chern-Simons approach to self-dual   
gravity in (2+1)-dimensions and quantisation of Poisson structure  }

\baselineskip 20 pt

\vspace{.2cm}

{ \bf Prince~K.~Osei } \\
African Institute for Mathematical Sciences (AIMS), Accra, Ghana\\
  Quantum Leap Africa (QLA), Kigali, Rwanda
 \\
{pkosei@aims.edu.gh}

\vspace{0.3cm}

\today
\baselineskip 16 pt

\end{center}

\begin{abstract}

 The (2+1)-dimensional  analog self-dual  gravity  which is obtained via spacetime dimension reduction of the  (3+1)-dimensional Holst action without reducing the internal gauge group is studied. A Chern-Simons formulation  for this theory is constructed 
 based on the gauge group $SL(2,\CC)_\RR\rcross \Rsix$  and maps
 the  3d complex self-dual dynamical  variable and connection to  
 6d real variables  which combines into a 12d Cartan connection.
 The Chern-Simons approach leads to a real analogue for the self-dual action based on a larger symmetry group.
 The quantization process follows the combinatorial quantization method outlined for Chern-Simons theory.
 In the combinatorial quantization of the phase space  %on $\RR \times \Sigma_{g,n},$ where $\Sigma_{g,n}$ is a genus $g$ surface with $n$ punctures,
 the Poisson structure governing the moduli space of flat connections %on $(SL(2,\CC)_\RR\rcross \Rsix)^{n+2g}$ 
 which emerges is obtained using 
the  classical $r$-matrix for the quantum double $D(SL(2,\CC)_\RR)$ viewed as the double of a double $ D(SL(2,\RR)\dcross AN(2))$. 
 This quantum double gives the structure for quantum symmetries within the quantum theory for the model.
\end{abstract}

%\centerline{PACS numbers: 04.20.Cv, 02.20.Qs, 02.40.-k}
%\maketitle

\section{Introduction and Motivation}

  In loop quantum gravity, the $SU(2)$ Ashteker-Barbero variables  were invented \cite{BAsh,Immirzi} to address concerns over the imposition of reality conditions in the quantization of the original formulation of complex Ashteker gravity \cite{{Ashtekar}}. The role of the Barbero-Immirzi parameter $\gamma$, whose physical significance is still  relatively open  has since been studied extensively \cite{AMN,BittrichRyan,MercuriRandono,Holst,MercuriTveras, MercuriQuinn,Mercuri2,TaverasYunes,BojowaldDas,Alexandrov,Mercuri,PerezRovelli,RovelliThieman,Marugan,FreidelMT,AGKY}. 
  In \cite{GeillerNouiH}, a three-dimensional Lorentzian gravity with a Barbero-Immirzi parameter containing an explicit Holst-like term was introduced.
  This model,  originally introduced in \cite{GN1} in the context of spin foam models to enable one to construct a complete spin form quantisation, was formulated  to understand the role of the Barbero-Immirzi parameter and the choice of connection in the construction of loop quantum gravity. See also  \cite{AGKY}.  The action is obtained via dimension reduction of the  (3+1)-dimensional Holst action but maintaining the same internal gauge group. In this dimensional  reduction, one impose invariance on the four-dimensional Holst action  along a given spatial direction to reduce the original version to an action for three-dimensional gravity containing a Barbero-Immirzi parameter.
\\

While the spin-foam models are formulated in terms of $SL(2,\CC)$ holonomies, the canonical quantisation program seem to have focused largely on the $SU(2)$ gauge group.
Starting from this action of three dimensional gravity studied in \cite{GN1,GeillerNouiH, AGKY}, 
we explore canonical quantisation of the self dual formulation of this model. 
More precisely, we are interested in how to quantise complex connections or  the implementation of the so called reality conditions arising when working with a complex connection.
To this end, we construct a Chern-Simons theory  based on the gauge group $SL(2,\CC)_\RR\rcross \Rsix$. A Chern-Simons theory on a three dimensional manifold depends on a gauge group and an invariant, non-degenerate symmetric, bilinear form on the Lie algebra of that gauge group.
Here, 
the dynamical variable, a $1$-form $E^i_\mu$ on the (2+1) space-time  with values in the 3d complex  Lie algebra  $\mathfrak{sl}(2,\RR)_\CC$   is view as  a 6d real valued $1$-form  $E^\alpha_\mu$ in  $\mathfrak{sl}(2,\CC)_\RR^*\cong \RR^6$.
The 3d complex self-dual %$\mathfrak{sl}(2,\CC)$ valued 
 connection $A^i_\mu$ is  viewed as a real 6d  $\mathfrak{sl}(2,\CC)_\RR$ valued connection $A^\alpha_\mu$. 
These are then combined into a  one-form $\calA$ on the space-time with values in the 12d Lie algebra $\mathfrak{sl}(2,\CC)_\RR\rcross \Rsix$ for the Chern-Simons action.
The Chern-Simons approach considered here leads to a real analogue of the self-dual action, but based on a larger symmetry group.
 This construction therefore provides a procedure for implementing some kind of reality condition on the the variables and action before quantization.
 \\

Given a three dimensional space-time manifold $M^3$ with topology $\RR\times \Sigma$, where $\Sigma$ is an orientable two-dimensional manifold that physically represents 'space',
the phase space of the Chern-Simons theory with gauge group $H$  on the manifold $M^3$ is  the  moduli space of flat $H$-connections modulo gauge transformations on the spacial surface $\Sigma$ \cite{Witten}.
Quantization can be obtained given by the application of the combinatorial quantisation programme  \cite{AGSI,AGSII,AS,Schroers,BNR,MeusburgerSchroers1, MeusburgerSchroers2,MN} to the Chern-Simons formulation of (2+1)-dimensional  gravity \cite{AT,Witten,Carlipbook}.
The discrete descriptions of the phase space underlying  the combinatorial quantisation formalism  and its application  to  Chern-Simons theory is originally  due to the original  work of Fock and Rosly \cite{FockRosly} and play a central role in the quantisation program.
\\
Here, we follow  the case of (non-compact and non-semisimple) semidirect product gauge groups of type $H=G\rcross \cg^*$ studied in \cite{MeusburgerSchroers2}, where the Poisson structure was discussed and  quantisation procedure developed. This allows one to consider the algebra of function $C^\infty (SL(2,\CC)_\RR)$.
\\

The Fock and Rosly’s description of the phase space depends on an oriented graph $\Gamma$ that is marked at each vertex and  represents the  oriented surface $\Sigma$. It also rely on a classical $r$-matrix for the Lie algebra corresponding to the gauge group that  is compatible with the Chern-Simons action.  Since closed oriented surfaces are classified by their fundamental groups,
 the moduli space of flat connections on the underlying surface $\Sigma$ can therefore be parametrised by a set of holonomies along closed paths that generate the fundamental group of $\Sigma$, modulo gauge transformations at the common starting and end points of those paths. 
In the combinatorial description of the phase space  on $\RR \times \Sigma_{g,n},$ where $\Sigma_{g,n}$ is a genus $g$ surface with $n$ punctures, the Poisson structure for the moduli space of flat  $SL(2,\CC)_\RR\rcross \Rsix$-
 connections 
 which emerge
   is defined  in terms of the  classical $r$-matrix for the quantum double $D(SL(2,\CC)_\RR)$ which is parametrised as the double of a double $ D(SL(2,\RR)\dcross AN(2))$. We describe how the quantum symmetries arising from the quantisation is determined by this quantum group.
   \\

This paper is organised as follows.
 We start in Sect.~2 by specifying some notations and  providing a  review  of the relevant Lie algebras for the construction. In particular we study the double cross sum Lie algebra $\mathfrak{h}=\mathfrak{sl}(2,\CC)_\RR\oplus \Rsix$  and provide a review of the left-and right-invariant vector fields of the  Lie group $SL(2,\CC)_\RR$ in terms of the coordinate functions.
  Sect.~3,  reviews the dimensional reduction of the  4d Holts action to the 3d analogand analyse the corresponding 3d self dual action which emerge in obtained the symmetry reduction.
   In   Sect.~4, we formulate a Chern-Simons theory for the self-dual variables based on the gauge group $SL(2,\CC)_\RR\rcross \Rsix.$
  Sect.~5 provide a description of the  
   Fock-Rosly  Poisson structure on $(SL(2,\CC)_\RR\rcross \Rsix)^{n+2g}$ which emerge in the combinatorial description of the phase space  of the Chern-Simons theory of Sect.~4 on $\RR \times \Sigma_{g,n},$ where $\Sigma_{g,n}$ is a genus $g$ surface with $n$ punctures.
  In Sect.~6, we discus quantisation and quantum symmetries.
  Sect.~7 contains  concluding remarks.

 \section{Conventions  and Lie algebras}
 
We use  $\mu, \nu, ...$ for space-time indices and $a, b, . . .$ for spatial part of the space-time indices. The indices $I , J, . . . $ for $\mathfrak{so}(3,1)$ or $\mathfrak{so}(4)$ in (3+1)-dimensions whiles $i, j, . . .$  denotes $\mathfrak{sl}(2,\RR)$ or $\mathfrak{su}(2)$ indices in (2+1)-dimensions.  
The metric is chosen to be $\eta_{IJ}=\eta^{IJ}$,  which  is, respectively,   the Euclidean metric diag$(1,1,1,1)$ or the Lorentzian metric diag$(-1,+1,+1,+1)$ in (3+1)-dimensions and $\eta_{ij}=\eta^{ij}$ which is diag$(1,1,1)$ or diag$(-1,+1,+1)$ in the Euclidean or Lorentzian  signatures respectively in (2+1)-dimensions.
The total antisymmetric tensor $\epsilon_{IJKL}$ is the Killing form on $\mathfrak{sl}(2,\CC)$ and $ \delta_{IJK} =(\eta_{IK}\eta_{JL}-\eta_{IL}\eta_{JK})/2$.
Any vector $v$ in $\RR^4$ with components $(v^0,v^i)$ so that the vector product between elements $\bu,\bv \in \RR^3$ is given by  the operation $(\bu\times \bv)^i = \epsilon_{ijk}u^jv^k$ where $\epsilon$ denotes the fully antisymmetric tensor in three
dimensions with the convention $\epsilon_{012}=\epsilon^{012}=1$ and the scalar product by $\bu\cdot \bv= \eta_{ij}u^iv^j.$  %Symmetrization and anti-symmetrization of indices are defined with  weight $1/2$.  
\\
 
We denote by $SL(2,\CC)$ the complex Lie group and by  $SL(2,\CC)_\RR$ the realification of  $SL(2,\CC)$. 
Following conventions in \cite{ Schroers,MeusburgerSchroers1, MeusburgerSchroers2}, we write elements in $H=SL(2,\CC)_\RR\rcross \Rsix$ as a pair $(u,\ba)$ so that group multiplication is given by
\bee
(u_1,\ba_1)\cdot (u_2,\ba_2) = (u_1\cdot u_2, \ba_1 + \Ad^*(u_1^{-1})\ba_2  ),
\eee
where $\Ad^*(g)$ is the algebraic dual of $\Ad(g)$ which is the coadjoint representation of $SL(2,\CC)_\RR$ on  $\Rsix$ with 
\bee 
\langle \Ad^*(g) \bj ,\xi \rangle =\langle \bj, \Ad(g)\xi\rangle, \quad \forall \bj\in \Rsix,\xi \in \mathfrak{sl}(2,\CC)_\RR, g\in SL(2,\CC)_\RR
  \eee 
so that the coadjoint action of $g\in SL(2,\CC)_\RR$ is given by $\Ad^*(g^{-1})$. We use the parametrisation 
\bee\label{coadjh}
(u,\ba )= (u, -\Ad^*(u^{-1})\bj  )\quad u\in SL(2,\CC)_\RR, \ba,\bj\in \Rsix.
\eee
\\

The generators of both the three-dimensional rotation
algebra  $\mathfrak{su}(2)$ and the three-dimensional Lorentz
algebra $\mathfrak{sl}(2,\RR)$ are denoted by $J_{i}$, with the distinction between the two cases again given by the context.
 In terms of these generators the Lie brackets
and Killing form are respectively 
\begin{equation}
\label{basicbracket}
\left[J_{i},J_{j}\right]=\epsilon_{ijk}J^{k}\quad \mbox{ and }\quad  \langle J_{i},J_{j}\rangle=\eta_{ij}.
\end{equation}
 The real  Lie algebra  of  $SL(2,\CC)_\RR$ denoted by 
 $\mathfrak{sl}(2,\CC)_\RR$ 
 has  relations
\bee
\label{sl2c}
[J_i,J_j]=\epsilon_{ijk}J^k, \quad  [J_i,K_j]=\epsilon_{ijk}K^k, \quad [K_i,K_j]=-\epsilon_{ijk}J^k.
\eee
Define the generator $\PT_i$ by  
\begin{equation}
 \PT_i=K_i+\epsilon_{ijk}n^jJ^k, \quad \bn^2<-1.
\label{Iwa decomp}
\end{equation}
Then the Lie brackets on $\mathfrak{sl}(2,\CC)_\RR$ take the form 
\begin{equation}
\left[J_{i},J_{j}\right]=\epsilon_{ijk}J^{k},\;\left[J_{i},\PT_{j}\right]=\epsilon_{ijk}\PT^{k}+n_jJ_i-\eta_{ij}(n^kJ_k),\;\;\left[\PT_{i},\PT_{j}\right]=n_i\PT_j-n_j\PT_i.\label{Iwa brackets}
\end{equation}
This is the Iwasawa decomposition of $\mathfrak{so}(3,1)\simeq\mathfrak{sl}(2,\CC)_\RR$
into a compact part $\mathfrak{su}(2)$ and  a non-compact part consisting of traceless, complex upper triangular matrices with real  diagonal with corresponding Lie group  $AN(2)$,   the group of  $2\times 2 $ matrices of the form
\begin{equation}
\label{an2para}
\begin{pmatrix}
e^{\alpha} & \xi+i\eta \\
0 & e^{-\alpha} \\
 \end{pmatrix},\quad \alpha ,\xi,\eta \in \RR.
 \end{equation}
 The notation refers the abelian and the nilpotent parts of this group, 
which is isomorphic to the semidirect product $\RR \rcross \RR^2.$
Explicitly, with choice $\bn=(1,0,0)$, we obtain 
\begin{equation}
 [\PT_0,\PT_a]=\PT_a, \quad [\PT_a,\PT_b]=0, \quad a,b=1,2
\end{equation}
which are familiar brackets for $\mathfrak{an}(2)$.
Alternative
 generators \begin{equation}                          
J_{i}^{\pm}=\frac{1}{2}\left(J_{i}\pm i K_{i}\right)\end{equation}
 of $\mathfrak{sl}(2,\CC)_\RR$ can be introduced in terms of which the Lie bracket takes the form
 \begin{equation}
\left[J_{i}^{\pm},J_{j}^{\pm}\right]=\epsilon_{ijk}J_{\pm}^{k},\quad \left[J_{i}^{\pm},J_{j}^{\mp}\right]=0.\label{alg of so(2,2)}
\end{equation}
The Lie algebra in this case is  a direct sum of two commuting copies of $\mathfrak{su}(2)$ or $\mathfrak{sl}(2,\RR)$. %For later use we also note that  with $J_i=J_i^++J_i^-$.
\\

 The Lie algebra of the Lie group $H$ is $\mathfrak{h}=\mathfrak{sl}(2,\CC)_\RR\oplus \Rsix$ and  generated by   $\calP_\alpha,\;\calJ_\alpha$  such that $\calJ_\alpha$ generates $\mathfrak{sl}(2,\CC)_\RR$ and $\calP_\alpha$ generates $\mathfrak{sl}(2,\CC)_\RR^*\cong \Rsix$  with $\alpha = 1, ..., 6$. These generators satisfy the relation
 \bee
 \label{genalgebra}
[\calJ_\alpha, \calJ_\beta]=f_{\alpha\beta}\;^{\gamma} \calJ_{\gamma}, \quad[\calJ_\alpha, \calP^\beta]=-f_{\alpha\gamma}\;^{\beta} \calP^{\gamma}\quad [\calP^\alpha,\calP^\beta]=0.
\eee
where  $f_{\alpha\beta}^\gamma$ are the structure constants for $\mathfrak{sl}(2,\CC)_\RR$. One can identify  $\calJ_\alpha $ with  the basis
 \begin{equation}\label{Jbasis} \calJ=\{J_0,J_1,J_2,K_0,K_1, K_2\}
 \end{equation}
 of $\mathfrak{sl}(2,\CC)_\RR$ to see that the first bracket in (\ref{genalgebra}) is that of  $\mathfrak{sl}(2,\CC)_\RR \simeq \mathfrak{so}(3,1)$  in \eqref{sl2c}
viewed  as the complexification of  $\mathfrak{su}\left(2\right)$
and  $\mathfrak{sl}\left(2,\mathbb{R}\right)$ for Euclidean
and Lorentzian signature, respectively.
The generators $\calP_\alpha$ of the $\mathfrak{sl}(2,\CC)_\RR^*$ viewed as $\Rsix=\RR^3\times\RR^3$ has basis 
\begin{equation}\label{Pbasis} \calP=\{P_0,P_1,P_2,\P_0,\P_1, \P_2\}
\end{equation}
satisfying
\bee
\label{sl2c*}
[P_i,P_j]=  [\P_i,\P_j]= [P_i,\P_j]=0.
\eee
The cross relation between (\ref{Jbasis}) and (\ref{Pbasis}) follows from (\ref{genalgebra}) as 
\bee
\label{crossrel}
[J_i,P_j]=-\epsilon_{ijk}P^k, \quad  [J_i,\P_j]=-\epsilon_{ijk}\P^k, \quad [K_i,P_j]=-\epsilon_{ijk}P^k,\quad [K_i,\P_j]=-\epsilon_{ijk}\P^k.
\eee
The Lie algebra $\mathfrak{sl}(2,\CC)_\RR\oplus \Rsix$ has a symmetric Ad-invariant  non-degenerate bilinear   form 
\bee
\label{form} \langle \calJ_\alpha,\calJ_\beta\rangle=0, \quad \langle \calP_\alpha,\calP_\beta\rangle=0,\quad  \langle \calJ_\alpha,\calP^\beta \rangle= \delta^\beta_\alpha.
\eee
In terms of the basis (\ref{Jbasis}) and (\ref{Pbasis}), we have 
\begin{align}
\label{pair} &\langle J_i,J_j\rangle=0, & &\langle K_i,K_j \rangle=0, & 
& \langle J_i,K_i \rangle =0,  \nonumber \\
&\langle P_i,P_j\rangle=0, & &\langle \P_i,\P_j \rangle=0, &
& \langle P_i,\P_i \rangle =0,  \nonumber \\
&\langle J_i,P_j\rangle=\eta_{ij}, & &\langle J_i,\P_j \rangle=\eta_{ij}, &
& \langle K_i,P_i \rangle =\eta_{ij},  \nonumber \\
&\langle K_i,\P_j\rangle =\eta_{ij}. & & &
&
\end{align}
\\

We denote by $\calJ_\alpha^L, \calJ_\alpha^R$ the left-and right-invariant vector fields on $SL(2,\CC)_\RR$ associated to the generators $\calJ_\alpha$ and defined by
\bee\label{vf}
\calJ_\alpha^RF (u) := \frac{d}{dt}|_{t=o}F(ue^{t\calJ_\alpha}),\quad \calJ_\alpha^LF (u) := \frac{d}{dt}|_{t=o}F(e^{-t\calJ_\alpha}u),%\; \forall u \in  SL(2,\CC)_\RR, F \in C^\infty(SL(2,\CC)_\RR).
\eee
for $ u \in  SL(2,\CC)_\RR$ and $ F \in C^\infty(SL(2,\CC)_\RR). $
The Lie group $SL(2,\CC)_\RR$  can be   viewed as   $ SL(2,\CC)_\RR=SU(2) \dcross AN(2) $ in Euclidean signature $SL(2,\CC)_\RR=SL(2,\RR)\dcross AN(2)$ in Lorentzian signature. 
We employ a parametrisation of this group that relies on decomposing it into an element
$u \in S U (2)$ or $SL(2,\RR)$  and an element $s \in AN (2)$ \cite{MeusburgerSchroers6,Meusburger3,OseiSchroers1}. The $ S U (2)$ or $SL(2,\RR)$ group element is parametrised by
\begin{equation}
u=p_3 + p^iJ_i, \qquad \frac{\bp^2}{4}+p_3^2=1,
\end{equation}
and the $AN(2)$ by 
\begin{equation}
v=q_3 + q^i\PT_i, \qquad q_3 =  \sqrt{1+(\bq\cdot\bn)^2/4},
\end{equation}
where $J_i$ and $\PT_i$ are defined in \eqref{basicbracket}  and (\ref{Iwa decomp}) respectively.  
We therefore write  an element $g \in SL(2,\CC)_\RR$ into elements parametrised above as 
\begin{equation}
g=(p_3 + p^iJ_i) \cdot (q_3 + q^i\PT_i)
\end{equation}
and  use the coordinate functions 
\begin{equation}
p^i : g\mapsto p^i, \quad  q^i : g\mapsto q^i.
\end{equation}
The left-and right-invariant vector fields generated by $J_i$ and $S_i$ in terms of the coordinate functions $p^i,q^i$ are given by \cite{MeusburgerSchroers6}
\bea \label{lrvectorfields}
J_i^Lp^i&=&-\eta^{ij}p_3+\frac{1}{2}\epsilon^{ijk}p_k,  \quad\quad\quad\quad\quad J_i^Lq^i=0, \nonumber\\
J^R_ip^i&=&\frac{q_3 - \frac{1}{2}\textbf{q} \cdot \textbf{n}  }{q_3 + \frac{1}{2}\textbf{q}  \cdot \textbf{n}  } (p_3\delta_i^j + \frac{1}{2}\epsilon_i^{jk}p_k )+ \frac{q_i}{q_3 + \frac{1}{2}\bq \cdot \bn } n^iq^j,  \quad J_i^Rq^i=\frac{q_3\epsilon_i^{jk}q_k-  \frac{1}{2}q^i\epsilon^{jkl}n_kq_l}{q_3 + \frac{1}{2}\bq \cdot \bn },\nonumber\\
\PT^L_ip^i&=&\frac{p_3}{2} p^j \epsilon_{ikl}p^kn^l+p_3^2n_ip^j+\frac{1}{4} \bp\cdot\bn p_ip^j- p_in^j,
 \nonumber\\
 \PT^L_iq^i&=&-\left((1-\frac{1}{2}\bp^2)\eta_{ik}+\frac{1}{2}p_ip_k-p_3 \epsilon_{ikl}p^l\right) \left((q_3 - \frac{1}{2}\bq\cdot \bn )\eta^{jk}+\frac{1}{2} n^kq^j \right),
 \nonumber\\
\PT^R_iq^i&=&(q_3 + \frac{1}{2}\bq \cdot \bn )\eta^{ij}-\frac{1}{2} n^iq^j, \quad \quad \PT_i^Rq^i=0.
\eea

%The Lie algebras $\mathfrak{sl}(2,\CC)_\RR$ is realised as the set of (split) quaternions over $\CC$ with vanishing unit component, 
% \begin{equation}
%\mathfrak{q}(\CC):=\{g\in\mathbb{H}_1(\CC)|\Pi(g)=0\},\end{equation}
%where $\Pi$ is a projector 
%$$ \Pi: \mathbb{H}_{1}(\CC)\rightarrow \mathbb{H}_{1},\quad \Pi: p_3+\bp\cdot\be\mapsto p_3.
%$$

\section{ Holst action in (2+1)-dimensions}

In this section, we analyse  the 3d analog  Holst action obtained via symmetry reduction.
This symmetry reduction involves  imposing invariance along a given spatial direction, which reduces the original four-dimensional Holst action to an action for three-dimensional gravity   with a Barbero-Immirzi parameter.

\subsection{Three dimension Holst action from 4d action via symmetry reduction}

In four-dimensional canonical loop quantum gravity, the first order action for general relativity is given by \cite{Holst} 
\begin{equation}
\label{Holst}
S_{\text{Holst}}[e,\omega]=\frac 1 4 \int_{M} \left(\frac{ 1}{ 2 }\epsilon_{IJKL} e^I\wedge e^J\wedge F^{KL} +\frac1 \gamma \delta_{IJKL} e^I\wedge e^J\wedge F^{KL}\right),
\end{equation}
where $\gamma$ is the analog of the Immirzi parameter. Here the dynamical variables are the tetrad $1$-form fields $e^I_\mu$ and the  $\mathfrak{sl}(2,\CC)$-valued connection $\omega_\mu^{IJ}$ with curvature 
%$F=d\omega +\frac 1 2 (\omega\wedge\omega).$ 
defined by 
\begin{equation}\label{curvature}
F^{IJ}_{\mu\nu}=\partial_\mu \omega^{IJ}_{\nu}-\partial_\nu\omega^{IJ}_\mu +\omega^{IL}_\mu \omega^{J}_{\nu L}-\omega^{IL}_\nu\omega^{J}_{\mu L}.
\end{equation}
One can perform a space-time reduction while maintaining the same  internal gauge group. 
Suppose the third spacial component $\mu=3$ is singled out, one can view   the four-dimensional space-time with topology $M^4 = M^3 \times \mathbb{I}$ where $M^3$ is a three-dimensional space-time, and $\mathbb{I}$  is a space-like segment with coordinates $x^3$.  Next, the following conditions are imposed 
\begin{equation} \partial_3=0, \quad \quad \quad\omega_3^{IJ}=0.
\end{equation}
These  conditions mean that the fields do not depend on the third spatial direction $x^3$ and that the parallel transport along $\mathbb{I}$ is trivial respectively. The covariant derivative of the fields along the direction $\mu = 3$ therefore vanishes. The four-dimensional Holst action (\ref{Holst}) then  reduces under these conditions to
%\begin{equation}
%\label{RedHolst}
%S_{\text{Holst}_{3d}}[e,\omega]=\int_\mathbb{I} dx^3 \int_{M^3} \textbf{d}^3x\epsilon^{\mu\nu\rho}\left(\frac{ 1}{ 2 }\epsilon_{IJKL} e^I_3 e^J_\mu F^{KL}_{\nu\rho} +\frac1 \gamma \delta_{IJKL} e^I_3 e^J_\mu F^{KL}_{\nu\rho} \right),
%\end{equation}
\begin{equation}
\label{HRaction}
S_{\text{Holst}_{3d}}[\chi,e,\omega]=\int_{M^3} \textbf{ d}^3x\epsilon^{\mu\nu\rho}\left(\frac{ 1}{ 2 }\epsilon_{IJKL} \chi^I e^J_{\mu} F^{KL}_{\nu\rho} +\frac1 \gamma \chi^Ie^J_\mu F_{\nu\rho IJ}\right).
\end{equation}
where $\mu= 0, 1, 2$ is now interpreted  as a three-dimensional space-time index, and $\textbf{d}^3x\epsilon^{\mu\nu\rho}$ is the local volume form on the three-dimensional space-time manifold and we have set  $\chi^I \equiv e^I_3$.
This is the action for three-dimensional Lorentzian gravity with a Barbero-Immirzi parameter. It was  introduced in   \cite{{GeillerNouiH,GN1}} to enable one to construct a complete spin form quantisation and studied in \cite{AGKY}
to understand role of the Barbero-Immirzi parameter and the choice of connection in the construction of loop quantum gravity.
An  interesting example of how dimensional reduction can be achieved via appropriate boundary conditions to lead a 4D Abelian gauge model to behave like  a 3D scalar system can be found in \cite{edery20093}.

\subsection{Lagrangian analysis  and symmetries}

Although the action (\ref{HRaction}) contains an extra degree of freedom in the variable $\chi$ and has an $SL(2,\CC)$ internal gauge group symmetry in Lorentzian signature or $SO(4)$ gauge group symmetry  the Euclidean picture,
one can show that it corresponds to three-dimensional gravity, and that similar to the four-dimensional Holst action, the Barbero-Immirzi parameter disappears once the torsion is vanishing. 
It is shown in \cite{AGKY} that
once the torsion-free condition is imposed, the action \eqref{HRaction}  reduces to the standard Einstein-Hilbert action.
% \begin{align}
%\label{Caction2}
%S_{\RR}[E,A]&=  \int_{M^3}E_{\alpha}\wedge \calR^{\alpha} . 
%\end{align}
We also refer to \cite{AGKY} for an account on the symmetries of the action  (\ref{HRaction})  such as an invariant under space-time diffeomorphisms,
 invariant under rescaling symmetry generated by a scalar and under
translational symmetries generated by a vector, etc.
%\[
%e^I_\mu\rightarrow \alpha e^I_\mu, \quad x^I \rightarrow \frac{ 1}{\alpha} x^I, \quad e^I_\mu\rightarrow  e^I_\mu + \beta_\mu x^I.
%\]
\\

 From \eqref{HRaction}, the dynamical variable of the three dimensional Holst action is now a $1$-form $E$ on the 3d space-time  with values in the Lie algebra  $\mathfrak{sl}(2,\CC)$  or $\mathfrak{so}(4)$ 
and given by
\bea
\label{Efield}
E^{IJ}_\mu=\epsilon^{IJ}\,_{KL}\chi^Ke^L_\mu,
\eea
where  %$e^I_\mu$ is a Lie-algebra valued $1$-form and 
$\chi^I$ is a $0$-form in $\RR^4$. 
The Hodge dual for the dynamical variable (\ref{Efield}) is 
\bea
\star E^{IJ} =\frac 1 2 \epsilon^{IJ}\,_{KL} E^{KL}.
\eea
 %One can then split this bivector into its self-dual and anti-self-dual components $E^\pm.$  
\\

Consider now the self-dual action for (\ref{HRaction}) which can be expressed as the complex-valued  
 action
\begin{align}
\label{SDaction}
S_\CC%=\int_{M^3}\textbf{ d}^3x\,\epsilon^{\mu\nu\rho}  E_\mu F_{\nu\rho} 
 = \int_{M^3}  E \wedge F[A],
\end{align}
in terms of the effective  self-dual variables 
\begin{equation} \label{SDvariables}
 E_\mu^j=(\chi^0e_\mu^j-\chi^je_\mu^0)  + i  \epsilon^{j}_{kl}\,\chi^k e^l_\mu%=E_{\mu}^{[0j]} \sign i  \epsilon^{j}_{kl}E_{\mu}^{\,{kl}}%=B_{\mu}^j  \sign i\epsilon^{j\,}_{kl}B_{\mu}^{\,{kl}}
 , \quad A_\mu^{j} = \omega_\mu^j\sign i \omega^{(0)j}_\mu. %=\omega_\mu^j\sign i\Gamma^{j}_\mu
\end{equation}
Here $E_\mu^j= E^{IJ}_\mu J^{+j}_{IJ} $ 
and $A$ denote  the self-dual component of $\omega$. We will refer to the action \eqref{SDaction} as the 3d analog self-dual gravity action.  One can show that the action \eqref{SDaction} can be obtained  from a decomposition of \eqref{HRaction} with $\gamma=i.$
We refer to Apendix \ref{analysis with gamma complex} for details.

\section{ Chern-Simons theory  for the self-dual variables  }
\label{Chern-S}

\label{ccssection}
Next, we write a Chern-Simons theory for the self-dual action \eqref{SDaction} described above by mapping the the complex self-dual variables to real ones.
The Chern-Simons theory on a three-dimensional manifold depends on a gauge group and an invariant, non-degenerate symmetric, bilinear form on the Lie algebra of that gauge group.

  \subsection{The Chern-Simons action for the self-dual variables}
  Consider a three-dimensional space-time manifold $M^3$ with topology $\RR\times \Sigma$, where $\Sigma$ is an orientable two-dimensional manifold that physically represents 'space'.
The gauge group is for the Chern-Simons theory is $$H=SL(2,\CC)_\RR\rcross \mathfrak{sl}(2,\CC)_\RR^*=SL(2,\CC)_\RR\rcross \Rsix,$$ a semidirect product of the six dimensional real Lie group $SL(2,\CC)_\RR$ and its dual  $\mathfrak{sl}(2,\CC)_\RR^*$, viewed as the abelian group $\Rsix$. 
The gauge field of the Chern-Simons theory considered here is locally a one-form $\calA$ on the space-time with values in the Lie algebra $\mathfrak{sl}(2,\CC)_\RR\rcross \Rsix$.
To obtain this, we first map the complex self-dual variables to  real ones by viewing the complex  $\mathfrak{sl}(2,\RR)_\CC$ or $\mathfrak{su}(2)_\CC$ variables $E_i, A^i$ in (\ref{SDvariables}) as real-valued  forms $E_\alpha, A^\alpha $ on $\Rsix$ and 
 $\mathfrak{sl}(2,\CC)_\RR$ respectively. 
In terms of the  basis (\ref{Jbasis}) and (\ref{Pbasis}), these take the form
\begin{equation}\label{decomposition1}
A=A^{\alpha}\calJ_{\alpha}=\omega_\mu^j J_j\sign \omega^{(0)j}_\mu K_j, \quad E= E_\alpha \calP^\alpha=(\chi^0e_\mu^j-\chi^je_\mu^0) P_j \sign \epsilon^{j}_{kl}\,\chi^k e^l_\mu
\P_j =B_{\mu}^j P_j \sign \calB^j \P_j
\end{equation}
where 
\begin{equation} \label{Cbasis}
K_j =i J_j,\quad  \quad\quad \P_j= iP_j.
\end{equation}

%\todo{What does the gauge group  mean $H=SL(2,\CC)_\RR\rcross \mathfrak{sl}(2,\CC)_\RR^*=SL(2,\CC)_\RR\rcross \Rsix$ mean in terms of model space-times?}

 The real variables   $A$ and $E$ then combine into a Cartan connection
\begin{equation}\label{Cconection}
\calA =A^{\alpha}\calJ_{\alpha}+E_\alpha \calP^\alpha
\end{equation}
for the Chern-Simons action.
The curvature of the connection $\calA$ is
 \bee \label{curvaturecalA}
\calF=d \calA+\calA\wedge\calA = F+T,
\eee
and combines the curvature for the $\mathfrak{sl}(2,\CC)_\RR$ valued spin connection 
\bee
F^\gamma =dA^\gamma+\frac{1}{2}f_{\alpha\beta}^{\gamma}A^\alpha\wedge A^\beta
\eee
and torsion 
\bee
T=(dE^{\gamma}+ f^{\alpha\beta\gamma}A_{\alpha}\wedge E_{\beta} )P_\gamma.
\eee
In terms of the decomposition \eqref{decomposition1}, these are given respectively by 
\begin{equation}\label{Rcurvature}
F= (d\omega^i+\frac 1 2 \epsilon^{i}_{\,jk}(\omega^j\wedge \omega^k -\omega^{(0)j} \wedge \omega^{(0)k}) )J_k+(d\omega^{(0)k}+\frac 1 2 \epsilon_{ijk}(\omega^j\wedge \omega^{(0)k} - \omega^{(0)j}\wedge \omega^k) )K_k\end{equation}
and
\begin{equation}\label{torsion of A}
T=(dB^i -\epsilon_{ij}\;^k(\omega^i +\omega^{(0)i})\wedge B^j) \,P_k +(d\calB^{i} -\epsilon_{ij}\;^k(\omega^i+\omega^{(0)i})\wedge \calB^{j}) \,\P_k.
\end{equation}
\\

The Chern-Simons action for the gauge field $\calA$ is 
\begin{equation}
\label{CSaction}
S_{CS}[\calA]=\frac 1 2\int_{M^3}\langle\calA \wedge d\calA\rangle+\frac{1}{3}\langle\calA\wedge[\calA,\calA]\rangle.
\end{equation}
Varying the action \eqref{CSaction} with respect to the gauge field $\calA$
amount to the flatness condition on the gauge field  
\beq \label{CSEinstein}
\calF =d\calA+\calA\wedge \calA= 0.\eeq
This is equivalent to  the condition of  vanishing torsion and the equations of motion
\bee
dE^{\gamma}+ f^{\alpha\beta\gamma}A_{\alpha}\wedge E_{\beta} =0, \quad \calR^\gamma =dA^\gamma+\frac{1}{2}f_{\alpha\beta}^{\gamma}A^\alpha\wedge A^\beta=0,
\eee
and becomes
\begin{equation}
dB^i -\epsilon_{ij}\;^k(\omega^i +\omega^{(0)i})\wedge B^j=0, \quad d\calB^{i} -\epsilon_{ij}\;^k(\omega^i+\omega^{(0)i})\wedge \calB^{j}=0
\end{equation}
and
\begin{equation}
d\omega^i+\frac 1 2 \epsilon^{i}_{\,jk}(\omega^j\wedge \omega^k -\omega^{(0)j} \wedge \omega^{(0)k}) =0,\quad d\omega^{(0)k}+\frac 1 2 \epsilon_{ijk}(\omega^j\wedge \omega^{(0)k} - \omega^{(0)j} \wedge \omega^k) =0
\end{equation}
on using the decomposition \eqref{decomposition1}.
\\

In general, for a Chern-Simons theory with gauge group $H$ on the 3d universe  $M^3$ of topology $\RR\times  \Sigma,$  
the  Chern-Simons gauge connection transforms according to the relation 
\bee
\mathcal{A}'=h \calA hi^{-1} + h d h^{-1}.
\eee 
where $h$ is an $H$-valued function on the surface $ \Sigma$. The invariance of the Chern-Simons action \eqref{CSaction} with respect to these transformations is a consequence of the Ad-invariance of the bilinear form $\langle\,,\,\rangle$. These correspond to infinitesimal diffeomorphism symmetries of gravity \cite{Witten}.
\\

The manifold with topology 
 $M^3=\RR\times  \Sigma$ enables one to decompose the gauge field is  with respect to the coordinate $x^0$ on $\RR$ according to 
 \bea \calA=\calA_0dx^0+\calA_ \Sigma, \eea
 where the spacial gauge field  $\calA_ \Sigma=\calA_adx^a$ is an $x^0$-dependent and Lie algebra valued $1-$form on the spatial surface $ \Sigma$ and $\calA_0$ is a Lie algebra valued function on $\RR\times  \Sigma$. The corresponding curvature is given by
 \beq
\calF=dx^0\wedge(\partial_0\calA_ \Sigma-d_ \Sigma\calA_0+[\calA_0,\calA_ \Sigma])+\calF_ \Sigma,
 \eeq
 where 
 \beq \calF_ \Sigma=d_ \Sigma \calA_ \Sigma+\calA_ \Sigma\wedge \calA_ \Sigma \eeq is a $2$-form on $ \Sigma$ and $d_ \Sigma$ denotes the exterior derivative on the surface $ \Sigma$.
In terms of this decomposition, the Chern-Simons action (\ref{CSaction}) become
\bee
S_{SC}[\calA_0,\calA_ \Sigma]=\int_\RR dx^0 \int_{ \Sigma} dx^2 \left( -\langle  \calA_\Sigma,\partial_0 \calA_\Sigma \rangle+ \langle A_0, \calF_\Sigma \rangle\right).
\eee
Thus the phase space variables are the components of the spacial gauge field $\calA_ \Sigma=\calA_adx^a$ and their canonical Poisson brackets are 
\begin{equation}
\{ \calA(x)^a_\ca,\calA(y)^b_\cb   \}=\epsilon^{ab}\delta^2(x-y)\langle X_\ca,X_\cb \rangle
\end{equation}
where the $X_\ca=\{ \calJ_\alpha, \calP_\beta \}_{ \alpha,\beta = 1,..., 6}$ are generators of the Lie algebra $\mathfrak{sl}(2,\CC)_\RR\rcross \Rsix$  and $\delta^2(x-y)$ is the Dirac delta function on $\Sigma$. 
The components $A_0$  of the gauge field act effectively as Lagrange multipliers and enforce the following equations of motions as primary constraints: 
\bee
\mathcal{F}_ \Sigma(x)=0.
\eee
Variation with respect to $A_\Sigma$ gives the evolution equation
$$ \partial_0\calA_ \Sigma= d_ \Sigma\calA_0+[ \calA_ \Sigma, \calA_0].
$$
These two equations are equivalent to the statement that the field strength $\calF$ vanishes which geometrically means the vanishing of curvature $F$ and  torsion $T$
\bee
F=0, \quad \quad T=0.
\eee
\\

To better understand  the relation  of (\ref{CSaction}) to \eqref{SDaction}
and the physical interpretation of the action (\ref{CSaction}), we use the decomposition (\ref{decomposition1}). After performing  integration by parts and dropping the boundary terms, the action (\ref{CSaction}) 
 can be expanded as  
 \begin{align}
\label{CSaction2}
S_{\RR}[E,A]&=  \int_{M^3}E_{\alpha}\wedge F^{\alpha}.
\end{align}
We refer to \eqref{CSaction2} as the real form of the complex self dual action  \eqref{SDaction}.
Thus the Chern-Simons approach considered here implements some kind of reality procedure, reproducing a real analogue \eqref{CSaction2} of the complex self-dual action  \eqref{SDaction}.
\\
To include the Barbero-Immirzi parameter in the Chern-Simons theory, one could  consider for example a more general Chern-Simons theory in the Euclidean setting
\begin{align}\label{SCgamma}
S^{\gamma}_{SC}=\left(\frac{\gamma+1}{\gamma}\right)S_{SC}[A]+ \left(\frac{\gamma-1}{\gamma}\right)\bar{S}_{SC}[\bar{A}],
\end{align}
which maps both the self-dual and anti-self-dual connections to reals $\mathfrak{so}(4)_\RR \rcross \Rsix$-valued variables. Analysis of this and how it compares to results in \cite{BonzomLivine,MSkappa} is deferred for future work.

\subsection{Physical phase space }

The phase space of the Chern-Simons theory with gauge group $H$ on a manifold $M^3$ is  the  moduli space of flat $H$-connections modulo gauge transformations on the spacial surface $\Sigma$ \cite{Witten}. 
It contains a canonical symplectic structure constructed by Atiyah and Bott  \cite{AtiyahBott,Atiyah} and  Goldman \cite{Goldman}.   This symplectic structure is determined by  the choice of a non-degenerate Ad-invariant symmetric bilinear form on the Lie algebra $\mathfrak{h}$ corresponding to  the Lie group $H$. We refer the reader to  \cite{Mess,RevMess,KS} for further details.
Sincet the classical solutions of the field equations or constraints in the Chern-Simons formulation described above are, flat $SL(2,\CC)_\RR\rcross \Rsix$-connections. We can therefore characterise the phase space of 3d gravity on the manifold $M^3$ in this formulation as the space of flat $SL(2,\CC)_\RR \rcross \Rsix$-connections  on $ \Sigma$ modulo gauge transformations.

\section{Phase space discretisation and Poison structure}
We now  discuss the discrete descriptions of the phase space underlying  the combinatorial quantisation formalism  and its  application  to the Chern-Simons theory under consideration. This construction is due to Fock and Rosly’s description \cite{FockRosly} and play a central role in the application of the combinatorial quantisation program to Chern-Simons theory \cite{AGSI,AGSII,AS,BNR,MeusburgerSchroers1, MeusburgerSchroers2,Schroers}.
We start by considering the formalism for a general Chern-Simons theory with gauge group $H$ and associated Lie algebra $\ch$.
The main components of  the Fock and Rosly’s description of the phase space are an oriented graph $\Gamma$ containing  a cilium at each vertex and a classical $r$-matrix for the gauge group that is compatible with the Chern-Simons action.

\subsection{Graph representation and phase space  variables }
 
 The oriented surface $\Sigma$ is represented by a directed graph $\Gamma$ embedded in the surface $\Sigma$ similar to ideas in that of lattice gauge theory.  
% The  connected components of $\Sigma \backslash\Gamma$ form the faces of $\Gamma$ and are required to be disjoint union of discs so that the graph is fine enough to resolve the topology of $\Sigma$.  
 Suppose  $V,E,F,$ denote the set of vertices, edges and faces of $\Gamma$ respectively. For each edge $e\in E$,  we denote by $s(e)$ and $t(e)$ the starting and target vertices respectively.
 The orientation of the surface implements  a cyclic ordering of the edge ends incident at each vertex $v\in V$. 
 Each vertex is also assigned by a  cilium that transforms this cyclic ordering into a linear one. 
 
 %The order of the edges is taken counterclockwise from the cilium and we write $e_1 < e_2$ if the edge end $e_1$ is of lower order than the edge end $e_2.$ 
  \begin{figure}\label{graphrep}
 
 \end{figure}
 The face $p$ of $\Gamma$ then correspond  to closed paths in $\Gamma$, up to cyclic permutations of paths that corresponds to different choices of their starting and endpoints.
 
%In the Chern-Simons formulation with gauge group $H$, the 
%$H$-valued phase space variables are obtained by integrating the Chern-Simons gauge field $\calA$ along paths on $\Sigma$.
 %Parametrising elements of the   $SL(2,\CC)_\RR\rcross \Rsix$ group as in (\ref{coadjh}), one assigns a $SL(2,\CC)_\RR$-element $u_\gamma$ and a vector $\bj\in \Rsix$ to each path $\gamma$
%  \bee\label{fpvariables}
%H_\gamma= (u_\gamma, -\Ad^*(u_\gamma^{-1})\bj_\gamma )=\text{Pexp}\int_\gamma \calA_\mu dx^\mu.
%\eee
%Reversing the orientation of the path $\gamma$ leads to variables which are related to the original ones as

 \subsection{ Moduli space of flat connections}
 
 Given the Lie group $H$, the graph connection on $H$ can be defined as the map
 \bee \calA:  E \rightarrow H,\quad e\mapsto h_e \eee
 which assigns an element of $H$ to each edge of the oriented graph $\Gamma$. The reversal of orientation corresponds to replacing $h_e$ with the inverse $h_e^{-1}$. 
 The $E$-fold product $H^{\times E}$ then gives the set $\calA(\Gamma)$ of graph connections on $\Gamma$.
 For each phase of $\Gamma$,  choose an associated path $p$ in $\Gamma$ such that the map
 \bee
 \mu_p: H^{| E|} \rightarrow H, \quad \mu_p(h_1, ..., h_n) =\prod_{e\in p} h_e^{\epsilon_e}
 \eee
determines the curvature of the graph connection at the phase $p$, where $\epsilon_e=\pm1$ if $e$ is in the orientation or opposite orientation of $p$. Tthis product must vanish for flat graph connection for paths  around each phase of the graph $\Gamma$. The space of flat graph connections is then given by 
\bea
{\calF}(\Gamma)=\bigcap_{p\in F} \mu^{-1}_p(1) \subseteq  H^{\times E}
\eea
and independent on of choices of paths.
Thus for a vertex $v\in V$, one defines the vertex action $ \la_v : H \times H^{ \times E} \rightarrow H^{\times E} $ by a group action of $H$ associated to the vertex $v\in \Gamma$. It acts only on the copies of $H$ that correspond to edges incident at $v$ and enforce graph gauge transformations.

The gauge transformations are described by group actions of $H$ associated to each vertex of the graph $\Gamma$.
 The action is define by left multiplication for incoming edges at $v$, right multiplication with the inverse for outgoing edges at $v$ and by conjugation for any loops at $v$.
%\bee
%\pi_e\left( h \la_v(h_1,...,h_n) \right) = \begin{cases} h \cdot h_e & v= t(e) \neq s(e)\\
%h_e\cdot h^{-1} & v= s(e) \neq t(e)\\
%h\cdot h_e\cdot h^{-1} & v= s(e) = t(e)\\
%h_e & v \notin \{t(e), s(e)\}
%\end{cases}
%\eee
%where  $\pi_e: H^{\times E}\rightarrow H, (h_1, ..., h_n)\mapsto h_e$ are the maps that project on the factor associated with the edge $e \in E$.
 %  Since the vertex actions  $\la_v$ and $\la_w$ commutes, one can combine them  to give the group action of $H^{\times V}$ on $H^{\times E}$  to obtain 
 The group of graph gauge transformations is $\calG(\Gamma)= H^{\times V}$ and  acts on the set $\calA(\Gamma)$ of graph connections.  The orbit space of this group action
 $$ \text{Hom}(\pi_1(\Sigma),H)/H = {\calF}(\Gamma)/\calG(\Gamma),
 $$
 then provides the moduli space of flat $H$-connection on $\Sigma$.
 We refer to \cite{MeusburgerPLG} for a recent  account.

 \subsection{Classical r-matrix and Fock-Rosly Poisson structure}
 
 The starting point of this construction is  a  description of the Poisson structure on an extended classical phase space  in terms of a classical $r$-matrix.  
 We refer the reader to  \cite{CP,Majidbook} for a  detailed account Lie bialgrbras and classical $r$-matrices.
A Lie bialgebra $(\cg,[\;,\;],\delta)$ is a Lie algebra $(\cg,[\;,\;])$ over a field $k$ equipped with a map $\delta:\cg\mapsto \cg\otimes  \cg$ (the
cocommutator, or cobracket) satisfying the following condition:
\begin{description}
 \item{(i)} $\delta:\cg\mapsto \cg\otimes \cg$ is a skew-symmetric linear map, i.e. $\delta:\cg\mapsto \wedge^2\cg$
\item{(ii)} $\delta$ satisfies the coJacobi identity 
 $
(\delta\otimes \text{id})\circ \delta(X)+\mbox{cyclic}=0, \quad\forall X\in \cg. 
$
%where $ Alt (a\otimes b\otimes c)=a\otimes b\otimes c+b\otimes c\otimes a+c\otimes a\otimes b.$
\item{(iii)} For all $X,Y\in \cg$,
 $ \delta([X,Y])=(\ad_X\otimes 1+1\otimes \ad_X)\delta(Y)-(\ad_Y\otimes 1+1\otimes \ad_Y)\delta(X).
 $
\end{description}
An element $r\in\wedge^{2}\cg$ is said to
be a coboundary structure of the Lie bialgebra $(\cg,[\;,\;],\delta)$ if
 $ \delta(X)=\ad_X(r)=[X\otimes1+1\otimes X,r]$. 
 \begin{definition}\label{rmatrix}
 For any Lie algebra $\ch$ with basis $\{X_\mathfrak{a}\},$  let $r=r^\mathfrak{ab}X_\mathfrak{a} \otimes Y_\mathfrak{b}\in \ch \otimes \ch $
 $r_{21}=\sigma(r)=r^{\ca\cb}Y_\cb\otimes X_\ca$ and set $r_{12}=r^{\ca\cb}X_\ca\otimes Y_{\cb}\otimes1,$ $r_{13}=r^{\ca\cb}X_\ca\otimes1\otimes Y_{\cb},$ $r_{23}=r^{\ca\cb}1\otimes X_\ca\otimes Y_{\cb},$ elements in $ \ch\otimes \ch\otimes \ch.$
 The element $r$ is a classical $r$-matrix for $\ch$  provided 
  it satisfies the classical Yang–Baxter equation (CYBE) 
 \begin{equation}\label{CYBE}
[[r,r]]=[r_{12,}r_{13}]+[r_{12,}r_{23}]+[r_{13,}r_{23}]=0.\end{equation}
and that  its symmetric component is ad-invarient
\bee  \ad_X(r+r_{21})=0,\quad \forall X\in \ch.
\eee
 \end{definition}
 A Lie bialgebra $(\cg,[\;,\;],\delta)$ is called quasitriangular if its cocommutator is of the form  $ \delta(X)=\ad_X(r)$ with a classical $r$-matrix.
 We refer to \cite{CP,Majidbook} for general background and \cite{OseiSchroers1,OseiSchroers3,Cracow,PKO1} for a detailed account  in the context of 3d gravity. 
The Fock and Rosly construction showed that if $H$ has a quasitriangular Poisson–Lie group structure, the Poisson structure on ${\calF}(\Gamma)/\calG(\Gamma)$
  can be described in terms of the qusitriangular structure. 
  This requires the following compatibility condition:
\begin{definition}\label{compatible}
% In the Fock-Rosly construction, 
A  classical $r$-matrix is said to be compatible with 
a Chern-Simons action if its symmetric part is equal to the Casimir associated  to the invariant, non-degenerate symmetric bilinear form used in the Chern-Simons action. 
 \end{definition}
The above Fock and Rosly's compatibility requirement  does not specify the  classical $r$-matrices uniquely. For a given Chern-Simons theory, one may find a family of compatible $r$-matrices. We refer to \cite{OseiSchroers1} for a detailed discussion on this and for a description of the general class of compatible $r$-matrices for a larger class of Chern-Simons theories based on the most  general symmetric bilinear  form on the isometry Lie algebras for 3d gravity.
\\

For the $SL(2,\CC)_\RR\rcross \Rsix$-Chern-Simons  being considered here, the appropriate  Casimir operator for the bilinear form (\ref{form}) is given by
\bee
K=\calJ_\alpha\tens \calP^\alpha +\calP^\alpha \tens \calJ_\alpha 
\eee
which in terms of the 
the basis  (\ref{sl2c}) and (\ref{sl2c*}) takes the form 
\bee
K=J_i\tens P^i+P^i\tens J_i+ S_i\tens \P^i+ \P^i  \tens S_i - \epsilon^{ijk}n_j \, (J_i\tens \P^i+ \P^i  \tens J_i ).
\eee
 The 
the classical $r$-matrix  compatible to the Chern-Simons action (\ref{CSaction}) is 
\bee
r=\calP^\alpha \tens \calJ_\alpha 
\eee
or 
\bee
r= P^i\tens J_i+\P^i\tens \PT_i - \epsilon^{ijk}n_j \,\P^i\tens J_i.
\eee
It has a coboundary Lie bialgebra structure with commutator (\ref{genalgebra}) and cocommutator %$\delta:\ch \mapsto \ch\otimes \ch$
\bee
\delta(\calJ_\alpha)=0,\quad  \delta(\calP^\alpha)= f_{\beta\gamma}\;^{\alpha} \; \calP^\beta \tens \calP^\gamma
\eee
 This amounts to equipping the Lie algebra $\mathfrak{h}=\mathfrak{sl}(2,\CC)_\RR\oplus \Rsix$ with the structure of a classical double $D(\mathfrak{sl}(2,\CC)_\RR)$. One  can also view this  as   $D(D(\mathfrak{sl}(2,\RR)))$ in the Lorentzian picture or  $D(D(\mathfrak{su}(2)))$ in the Euclidean picture.
  
 Denote now by $X_\ca^{L_e}$ and  $X_\ca^{R_e}$ the left and right invariant vector fields on the 
copy of $H$ for each edge $e \in E,$ whose action on functions $F \in C^\infty(H^{\times E})$ is given by
\begin{align} \label{edgevfs}
X_\ca^{L_e}\cdot F (h_1,...,h_n) &:= \frac{d}{dt}|_{t=o}F(h_1,...,e^{-tX_\ca}\cdot h_e,...,h_n),  \nonumber\\
X_\ca^{R_e}\cdot F (h_1,...,h_n) &:= \frac{d}{dt}|_{t=o}F(h_1,...,h_e\cdot e^{tX_\ca},...,h_n).
\end{align}
The Fock and Rosly’s Poisson structure is expressed  in terms of a Poisson bivector  
\bee
\{F,G\}=(dF\tens dG)(B_{FR})\quad \forall F,G\in \calC^\infty(H)
\eee
 where $B_{FR}$ is obtained  by the following Theorem.
 \begin{theorem}\label{maintheorem} Fock-Rosly Poisson Structure 
 \cite{FR1,FR2,FockRosly}
 
 Let $\Gamma$ a directed graph embedded into an orientable surface $\Sigma$ such that $\Sigma\backslash \Gamma$ is a disjoint union of discs  and $H$ be a quasitriangular Poisson-Lie group. To each vertex $v \in V$, allocate a classical $r$-matrix $r(v) = r^{\ca\cb}(v)T_\ca\tens T_\cb$, with symmetric parts being non-degenerate and match for all vertices $v \in V.$
 \begin{enumerate}
 \item[]
 Then the Poisson bivector given by
  \begin{align}
\label{bivector}
B = &  \sum_{v\in V} r_{(\ca)}^{\ca\cb}(v) \left(   \sum_{s(e)=v} T_\ca^{R_e}\wedge T_\ca^{R_e}     +  \sum_{t(e)=v}  T_\ca^{L_e}\wedge T_\cb^{L_e}                \right)  \nonumber \\ 
+ &  \sum_{v\in V} r^{\ca\cb}(v) \left(   \sum_{v=t(e)=t(p)} T_\ca^{L_e}\wedge T_\cb^{Lp}   +  \sum_{v=t(e)=s(p)}  T_\ca^{L_e}\wedge T_\cb^{R_p}     +  \sum_{v=s(e)=t(p)}  T_\ca^{R_e}\wedge T_\cb^{Lp}                \right)  \nonumber \\ 
+&  \sum_{v=s(e)=t(p)}  T_\ca^{R_e}\wedge T_\cb^{R_p}  
\end{align}
defines a Poisson $H^{\times V}$-space structure on  $H^{\times E}$   for the  $H^{\times V}$-action from. Here $r^{\ca\cb}(v)$ denotes  the components of the classical $r$-matrices associated to the vertex of $\Gamma$ and satisfying the compatibility conditions in Definition \ref{compatible}.
 
 \end{enumerate}
 
 \end{theorem}
 
In the Theorem above, the Poisson brackets of  $H^{\times V}$-invariant functions on  $H^{\times E}$   are independent of the choice of the
marking that determine the cyclic orientation of the graph but dependent only on the  symmetric component common to all $r$-matrices $r_{\ca\cb}(v)$.
For further details, we refer to  \cite{AlexseevMalkin, Spies, Kitaev, MeusburgerPLG}.

 \subsection{ The algebra for $SL(2,\CC)_\RR\rcross \Rsix$ on %a genus a $g$ surface $\Sigma_{n,g}$  with $n$ punctures.
 the space of holonomies}
 
 For the Chern-Simons theory with compact and semisimple gauge groups, the Poisson structure on the space of holonomies was constructed  by Alekseev, Grosse and Schomerus \cite{AGSI, AGSI,AS} and quantised via their formalism of combinatorial quantisation of Chern-Simons gauge theories.
In the following, we apply the construction in the previous section to the gauge group $H=SL(2,\CC)_\RR\rcross \mathfrak{sl}(2,\CC)_\RR^*$ which arise in the Chern-Simons formulation of for the  self-dual gravity model. 
 In particular, instead of using the $r$-matrix and the left and right invariant vector field on the various copies of  $SL(2,\CC)_\RR\rcross \Rsix$ to construct the algebra of function on $C^\infty (SL(2,\CC)_\RR\rcross \Rsix)$,  
 we follow aspects of the case of (non-compact and non-semisimple) semidirect product gauge groups of type $H=G\rcross \cg^*$ studied in \cite{MeusburgerSchroers2}, where the Poisson structure was discussed and  quantisation procedure developed and allow one to consider the algebra of function $C^\infty (SL(2,\CC)_\RR)$.
 \\
 
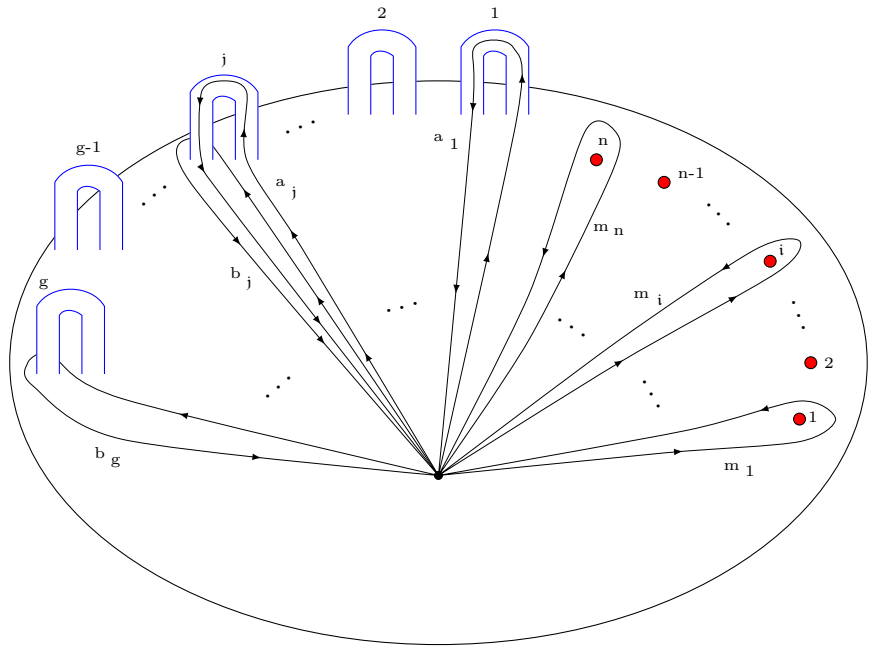
\begin{figure}[h]
 \begin{center}
\begin{tikzpicture}[scale=1.5]

    % \foreach \i in {0,...,\X}{
    %     \foreach \j in {1,...,\Y}{
    %         \draw[gray!40] (\i*\q, \j*\q) -- (\i*\q + \q, \j*\q);
    %     }
    % }

    % \foreach \i in {1,...,\X}{
    %     \foreach \j in {0,...,\Y}{
    %         \draw[gray!40] (\i*\q, \j*\q) -- (\i*\q, \j*\q + \q);
    %     }
    % }

    % \foreach \i in {0,...,\X}{
    %     \node[below] at (\i*\q, 0) {\tiny\i};
    % }

    % \foreach \j in {1,...,\Y}{
    %     \node[left] at (0, \j*\q) {\tiny\j};
    % }

    % \draw[color=gray] (17*\q, 8.5*\q) circle [x radius=3.3cm, y radius=2.0cm];

    \draw[color=black] (17*\q, 8.5*\q) circle [x radius=3.8cm, y radius=2.5cm];

    \draw[draw=black, fill=black] (17*\q, 6*\q) circle (1pt); %pivot

    %puncture 1
    \puncture{25}{7.25};
    \node at (25.30*\q, 7.30*\q) {\tiny 1};

    %path m_1
    \begin{scope}[very thin, decoration={markings, 
        mark=at position 0.3 with {\arrow{latex}},
        mark=at position 0.6 with {\arrow{latex}}}]
        \draw[postaction={decorate}] plot[smooth]
        coordinates {
            (17*\q, 6*\q) (22*\q, 6.5*\q) (25*\q, 6.75*\q) (25.8*\q, 7.25*\q) (25*\q, 7.65*\q) (22.5*\q, 7*\q) (17*\q, 6*\q)
        };
    \end{scope}

    %puncture 1
    \puncture{25.25}{8.5}
    \node at (25.65*\q, 8.5*\q) {\tiny 2};

    \node[rotate=-70] at (25*\q, 9.5*\q) {\ldots};

    %puncture i
    \puncture{24.35}{10.75};
    \node at (24.6*\q, 11*\q) {\tiny $\mathfrak{i}$};

    %path m_i
    \begin{scope}[very thin, decoration={markings,
        mark=at position 0.25 with {\arrow{latex}},
        mark=at position 0.4 with {\arrow{latex}},
        mark=at position 0.6 with {\arrow{latex}}}]
        \draw[postaction={decorate}] plot[smooth]
        coordinates {
            (17*\q, 6*\q) (21*\q, 8.5*\q) (24.5*\q, 10.5*\q) (25*\q, 11.2*\q) (24*\q, 11*\q) (21*\q, 9*\q) (17*\q, 6*\q)
        };
    \end{scope}

    \node[rotate=-45] at (23.25*\q, 11.75*\q) {\ldots};

    %puncture n-1
    \puncture{22}{12.5};
    \node at (22.6*\q, 12.7*\q) {\tiny n-1};
    
    %puncture n
    \puncture{20.5}{13}
    \node at (20.65*\q, 13.40*\q) {\tiny n};

    %path m_n
    \begin{scope}[very thin, decoration={markings, 
        mark=at position 0.3 with {\arrow{latex}},
        mark=at position 0.7 with {\arrow{latex}}}]
        \draw[postaction={decorate}] plot[smooth]
        coordinates {
            (17*\q, 6*\q) (19*\q, 9*\q) (20.75*\q, 12.5*\q) (21*\q, 13.5*\q) (20.26*\q, 13.65*\q) (19*\q, 10*\q) (17*\q, 6*\q)
        };
    \end{scope}

    %handle 1
    \node at (18.25*\q, 16.25*\q) {\tiny 1};
    \handle{17.5}{14};

    %path a_1
    \begin{scope}[very thin, decoration={markings, 
        mark=at position 0.25 with {\arrow{latex}},
        mark=at position 0.45 with {\arrow{latex}},
        mark=at position 0.6 with {\arrow{latex}},
        mark=at position 0.8 with {\arrow{latex}}}]
        \draw[postaction={decorate}] plot[smooth]
        coordinates {
            (17*\q, 6*\q) (18.75*\q, 14*\q) (18.65*\q, 15.5*\q) (17.8*\q, 15.5*\q) (17.75*\q, 14*\q) (17*\q, 6*\q)
        };
    \end{scope}
    
    % handle 2
    \node at (15.75*\q, 16.25*\q) {\tiny 2};
    \handle{15}{14};

    %path b_j
    \begin{scope}[very thin, decoration={markings, 
        mark=at position 0.25 with {\arrow{latex}},
        mark=at position 0.4 with {\arrow{latex}},
        mark=at position 0.65 with {\arrow{latex}},
        mark=at position 0.8 with {\arrow{latex}}}]
        \draw[postaction={decorate}] plot[smooth]
        coordinates {
            (17*\q, 6*\q) (12.4*\q, 12.75*\q) (11.7*\q, 13.5*\q) (11.5*\q, 12.5*\q) (17*\q, 6*\q)};
    \end{scope}

    %handle j
    \node at (12.25*\q, 15.25*\q) {\tiny $\mathfrak{j}$};
    \handle{11.5}{13};

    %path a_j
    \begin{scope}[very thin, decoration={markings, 
        mark=at position 0.15 with {\arrow{latex}},
        mark=at position 0.3 with {\arrow{latex}},
        mark=at position 0.42 with {\arrow{latex}},
        mark=at position 0.53 with {\arrow{latex}},
        mark=at position 0.6 with {\arrow{latex}},
        mark=at position 0.8 with {\arrow{latex}}}]
        \draw[postaction={decorate}] plot[smooth]
        coordinates {
            (17*\q, 6*\q) (14*\q, 11*\q) (12.75*\q, 13*\q) (12.75*\q, 14.5*\q) (12.25*\q, 14.75*\q) (11.75*\q, 14.5*\q) (11.75*\q, 13*\q) (12.35*\q, 12*\q) (17*\q, 6*\q)};
    \end{scope}

    % handle g-1
    \node at (9.25*\q, 13.25*\q) {\tiny g-1};
    \handle{8.5}{11};

    %path b_g
    \begin{scope}[very thin, decoration={markings, 
        mark=at position 0.3 with {\arrow{latex}},
        mark=at position 0.8 with {\arrow{latex}}}]
        \draw[postaction={decorate}] plot[smooth]
        coordinates {
            (17*\q, 6*\q) (10*\q, 7.7*\q) (8.5*\q, 8.6*\q) (8.25*\q, 8.75*\q) (7.9*\q, 8.5*\q) (8*\q, 8*\q) (10*\q, 6.8*\q) (17*\q, 6*\q)
        };
    \end{scope}

    % handle g
    \node at (8.25*\q, 10.25*\q) {\tiny g};
    \handle{8.1}{8.25};

    \node[rotate=22] at (14*\q, 13.75*\q) {\ldots};

    \node[rotate=40] at (10.75*\q, 12.25*\q) {\ldots};

    \node[rotate=40] at (13.5*\q, 8*\q) {\ldots};

    \node[rotate=15] at (16.25*\q, 9.75*\q) {\ldots};

    \node[rotate=-32] at (20*\q, 9.25*\q) {\ldots};

    \node[rotate=-60] at (21.75*\q, 7.75*\q) {\ldots};

    \node at (9.5*\q, 6.5*\q) {\tiny b};
    \node at (9.85*\q, 6.30*\q) {\tiny g};

    \node at (12.5*\q, 10.5*\q) {\tiny b};
    \node at (12.8*\q, 10.30*\q) {\tiny $\mathfrak{j}$};

    \node at (13.5*\q, 12.5*\q) {\tiny a};
    \node at (13.8*\q, 12.30*\q) {\tiny $\mathfrak{j}$};

    \node at (17*\q, 13.5*\q) {\tiny a};
    \node at (17.35*\q, 13.35*\q) {\tiny 1};

    \node at (20.6*\q, 11.5*\q) {\tiny m};
    \node at (21*\q, 11.40*\q) {\tiny n};

    \node at (21.5*\q, 10*\q) {\tiny m};
    \node at (21.9*\q, 9.9*\q) {\tiny $\mathfrak{i}$};

    \node at (23.5*\q, 6.2*\q) {\tiny m};
    \node at (23.9*\q, 6.1*\q) {\tiny 1};
    
\end{tikzpicture}
\caption{Generators of the fundamental group of a compact  surface $\Sigma_{n,g}$  with $n$ punctures }
\label{Surface} 
\end{center}
 \end{figure}

 The starting point in the phase space description is the recognition that flat connections on a manifold are characterised by their holonomies along non-contractible paths. Since closed oriented surface are classified by their fundamental groups,
 the moduli space of flat connections on the underlying surface $\Sigma$ can therefore be parametrised by a set of holonomies along closed paths which generate the fundamental group of $\Sigma$, modulo gauge transformations at the common starting and end points of those paths. 
For  the fundamental group $\pi_1(\Sigma_{n,g})$  of a genus $g$ surface 
 $\Sigma_{n,g}$  with $n$ punctures 
 are  generated by the equivalence classes loops $m_\ci$ $\ci = 1, . . . , n,$ around the punctures and two curves $a_\cj, b_\cj, $ $\cj = 1,...,g,$ for each handle, shown in Fig. \ref{Surface}. 

They are subject to the relation
\bea
[b_g,a_g^{-1}]\cdot ...\cdot [b_1,a_1^{-1}]\cdot m_n\cdot ...\cdot m_1=1, \quad \text{with} \quad [b_\ci,a_\ci^{-1}]=b_\ci a_\ci^{-1}b_\ci^{-1}a_\ci
\eea
The holonomies of the curves $A_\cj = \text{Hol}(a_\ci ), B_\cj= \text{Hol}(b_\cj )$ associated to each handle are general elements of the gauge group 
$SL(2,\CC)_\RR\rcross  \Rsix$. However the holonomies $M_\ci= \text{Hol}(m_\ci)$ around the punctures are contained in fixed $SL(2,\CC)_\RR\rcross \Rsix$-conjugacy classes
\bea \label{cclass}
\calC_{\mu_\ci s_\ci}= \{ (v,\bx)\cdot (g_\mu,-\bs)\cdot  (v,\bx)^{-1}\}|
( v,\bx)\in SL(2,\CC)_\RR\rcross \Rsix,
\eea
where $\mu_\ci$ label $SL(2,\CC)_\RR$-congugacy classes and $s_\ci$ represents co-adjoint orbits of corresponding   stabiliser Lie algebra.
The space  $\tilde{\calA}_{g,n}$  of graph connections  is given by 
\bea
\tilde{\calA}_{g,n}&=& \{(M_1,...,M_n, A_1,B_1,...,A_g,B_g)\in \calC_{\mu_1s_1}\times ...\times \calC_{\mu_ns_n}\times (SL(2,\CC)_\RR\rcross \Rsix )^{2n} | \nonumber\\
&&[A_g,B_g^{-1}]\cdot ...\cdot [A_1,B_1^{-1}]\cdot M_n\cdot ...\cdot M_1=1  \}
\eea
The moduli space $\mathcal{M}_{g,n}$ of flat $SL(2,\CC)_\RR\rcross \Rsix$-connections on a surface $\Sigma_{g,n}$ is then obtained from this extended phase space by dividing out conjugation of all holonomies by the group  $SL(2,\CC)_\RR\rcross \Rsix$
\bee
\mathcal{M}_{g,n}= \tilde{\calA}_{g,n}/\sim
\eee
where $\sim $ denotes silmultaneous conjugation.
In the application to $SL(2,\CC)_\RR\rcross \Rsix$, the holonomies are parametrised via (\ref{coadjh}) as
$X =\text{Hol}(x) =(u_X,-\Ad^*(u_X^{-1})\bj^X) \text{ for } $ 
\\$X \in \{M_1,...,M_n,A_1,B_1,...,A_g,B_g\}. $ The 
vectors $\bj^X$ in (\ref{coadjh}) are expanded as $\bj^X = j_\beta^X\calP^\beta$ and the same symbol denote the coordinate functions
$j_\alpha: (u, -\Ad^*(u^{-1})\bj  )\mapsto j_\alpha,$
$j_\alpha \in  C^\infty(SL(2,\CC)_\RR).$ 
Thus instead  of $C^\infty(SL(2,\CC)_\RR\rcross \Rsix)$ we considers the algebra generated by the functions in $C^\infty(SL(2,\CC)_\RR$ together with these maps $j_\alpha$.  We refer to \cite{MeusburgerSchroers2}  for more general account.

 \begin{definition}\label{flower} (Fock-Rosly algebra for the gauge group  $SL(2,\CC)_\RR\rcross\Rsix$)
 
 The  algebra $\bar{\calF}$ for gauge group $SL(2,\CC)_\RR\rcross \Rsix$ on a genus $g$ surface $\Sigma_{g,n}$ with $n$ punctures is the commutative Poisson algebra
\bea\label{flowereqn}
\bar{\calF} = S\left( \bigoplus_{\ck=1}^{n+2g}\mathfrak{sl}(2,\CC)_\RR   \right)\tens \calC^\infty(SL(2,\CC)_\RR)
\eea
where $S\left( \bigoplus_{\ck=1}^{n+2g}\mathfrak{sl}(2,\CC)_\RR   \right)$ is the symmetric envelope of the real Lie algebra
$\bigoplus_{\ck=1}^{n+2g}\mathfrak{sl}(2,\CC)_\RR $, i.e.  the  polynomials with real coefficients on the vector space $\bigoplus_{\ck=1}^{n+2g}\Rsix. $
 In terms of a fixed basis $\mathcal{B} =   j_\alpha^{M_\ci},  j_\alpha^{A_\ck},  j_\alpha^{B_\ck}, \ci=1, ..., n\;\ck=1, ..., g, \; \alpha=1, ...,6\}$,
 its Poisson structure is given by 
 \begin{align}
\label{floweralgebra}
\{ j^X_i ,j_j^X \}  = &  -\epsilon_{ij}\;^{k}   j_k^X     \nonumber \\ 
\{j^X_{i},\pt^X_{j}\}=&-\epsilon_{ij}\;^{k}\pt^X_{k}+n_jj^X_i+\eta_{ij}(n^kj^X_k),\nonumber \\ 
\{\pt^X_{i},\pt^X_{j}\}=&-n_i\pt^X_j+n_j\pt^X_i \nonumber \\ 
\nonumber \\
\{ j^X_i ,j_j^Y \}  = &  -\epsilon_{lj}\;^{k}   j_k^Y  (\delta_i^l - \Ad^*(u_X)_i\,^l)    \nonumber \\ 
\{j^X_{i},\pt^Y_{j}\}=&(-\epsilon_{ij}\;^{k}\pt^Y_{k}+n_jj^Y_i+\eta_{ij}(n^kj^Y_k)) )(\delta_i^l - \Ad^*(u_X)_i\,^l) ,\nonumber \\ 
\{\pt^X_{i},\pt^Y_{j}\}=&(-n_i\pt^Y_j+n_j\pt^Y_i) (\delta_i^l - \Ad^*(u_X)_i\,^l)   \quad\quad\quad\quad  \quad\quad  \forall X,Y\in \{ M_1,..., B_g     \}, X<Y    \nonumber \\ 
 \nonumber \\ 
\{ j^{A_\ci}_i ,j_j^{B_\ci} \}  = &  -\epsilon_{ij}\;^{k}   j_k^{B_\ci}     \nonumber \\ 
\{j^{A_\ci}_{i},\pt^{B_\ci}_{j}\}=&-\epsilon_{ij}\;^{k}\pt^{B_\ci}_{k}+n_jj^{B_I}_i+\eta_{ij}(n^kj^{B_\ci}_k),\nonumber \\ 
\{\pt^{A_\ci}_{i},\pt^{B_\ci}_{j}\}=&-n_i\pt^{B_\ci}_j+n_j\pt^{B_\ci}_i  \quad\quad \quad\quad  \quad\quad  \quad\quad \forall \ci=1, ..., g  \nonumber \\ 
\nonumber \\
\{ j_i^{M_\ci} ,  F\}  = &  - (J_i^{R_{M_\ci}} +J_i^{L_{M_\ci}} ) F-  (\delta_i^k - \Ad^*(u_{M_\ci})_i\,^k)  \left( \sum_{Y>M_\ci} (J_k^{R_Y} + J_k^{L_Y})F   \right)    \nonumber \\ 
\{ s_i^{M_\ci} ,  F\}  = &  - (S_i^{R_{M_\ci}} +S_i^{L_{M_\ci}} ) F-  (\delta_i^k - \Ad^*(u_{M_\ci})_i\,^k)  \left( \sum_{Y>M_\ci} (S_k^{R_Y} + S_k^{L_Y})F   \right)    \nonumber \\ 
\{ j_i^{A_\ck} ,  F\}  = &  - (J_i^{R_{A_\ck}} +J_i^{L_{A_\ck}} ) F- (J_i^{R_{B_\ck}}+J_i^{L_{B_\ck}})F - \Ad^*(u^{-1}_{B_\ck}u_{A_\ci})_i\,^j J_j^{R_{B_\ck}} 
 \nonumber \\
& -  (\delta_i^j  - \Ad^*(u_{A_i})_i\,^j)  \left( \sum_{Y>A_\ci} (J_\beta^{R_Y} + J_\beta^{L_Y})F   \right) \nonumber \\
\{ s_i^{A_\ck} ,  F\}  = &  - (S_i^{R_{A_\ck}} +S_i^{L_{A_\ck}} ) F- (S_i^{R_{B_\ck}}+S_i^{L_{B_\ck}})F - \Ad^*(u^{-1}_{B_\ck}u_{A_\ci})_i\,^j S_j^{R_{B_\ck}}   \nonumber \\
&- (\delta_i^j- \Ad^*(u_{A_\ci})_i\,^j)  \left( \sum_{Y>A_\ci} (S_\beta^{R_Y} + S_\beta^{L_Y})F   \right)  \nonumber \\
\{ j_i^{B_\ck} , F\}  = & - J_i^{L^{A_\ci}} F- (J_i ^{R_{B_\ci}}+J_i^{L_{B_\ci}})-  (\delta_i^k - \Ad^*(u_{B_\ci})_i\,^k)  \left( \sum_{Y>B_\ci} (J_k^{R_Y} + J_k^{L_Y})F   \right)
 \nonumber \\
 \{ s_i^{B_\ck} , F\}  = & - S_i^{L^{A_\ci}} F- (S_i ^{R_{B_\ci}}+S_i^{L_{B_\ci}})-  (\delta_i^k - \Ad^*(u_{B_\ci})_i\,^k)  \left( \sum_{Y>B_\ci} (S_k^{R_Y} + S_k^{L_Y})F   \right),
\end{align}
where $F\in \calC^\infty(SL(2,\CC)_\RR)$ are the coordinate functions $\{p^i,q^i\}$, $M_1< ... <M_n<A_1,$ \\
$ B_1< ... < A_g, B_g$ and $J_i^{L_X}, J_i^{R_X}, S_i^{L_X}, S_i^{R_X}$ denote the left- and right invariant vector fields  \eqref{lrvectorfields}  on $SL(2,\CC)_\RR)$.
 \end{definition}

The symplectic leaves of the Poisson manifold $(SL(2,\CC)_\RR\rcross \Rsix)^{n+2g}$ with bracket (\ref{floweralgebra}) are of
the form 
$ \calC_{\mu_1s_1}\times ...\times \calC_{\mu_ns_n}\times  T^*(SL(2,\CC)_\RR)^{2g},$
 where $\calC_{\mu_\ci s_\ci}$ denote $SL(2,\CC)_\RR\rcross \Rsix$-conjugacy classes as
in (\ref{cclass}).
%Note that these conjugacy classes arise as the symplectic leaves of the Poisson structure (\ref{floweralgebra}) on $(SL(2,\CC)_\RR\rcross \Rsix)^{n+2g}$. 
The Poisson structure on the symplectic leaves is given by a symplectic potential $\Theta$ on $(SL(2,\CC)_\RR\rcross \mathfrak{sl}(2,\CC)_\RR^*)^{n+2g}$.  This potential can be expressed in terms of the holonomies. See  \cite{MeusburgerSchroers2} for the general construction.

 \section{Quantisation and quantum symmetries  }%and the quantum double \\$D(SL(2,\CC)_\RR)$ } 
The task of quantising the flower algebra is to promote the discrete graph variables to an algebra of operators and to determine the irreducible representations which determine the space of quantum states.  This amounts to the quantisation of the symplectic structure on the cotangent bundle $T^*SL(2,\CC)_\RR$ of the group $SL(2,\CC)_\RR$ and the
dual Poisson structure $(SL(2,\CC)_\RR\rcross \Rsix)^*$.  Quantisation of Poisson algebras associated to each puncture and handle are combined to construct the quantum algebra of $SL(2,\CC)_\RR\rcross \Rsix$ and its irreducible representation.
% and then applying this to the quantization of the flower alagebra Both of these Poisson structures are relatively simple and special cases of the following general situation.

\subsection{Quantum algebra and representations  }
The quantum algebra for the flower algebra given in  (\ref{flowereqn}) is the associative algebra \cite{MeusburgerSchroers2}
\bea\label{qflowereqn}
\hat{\calF} = U\left( \bigoplus_{k=1}^{n+2g}\mathfrak{sl}(2,\CC)_\RR   \right)\hat{\tens}\; \calC^\infty\left(SL(2,\CC)_\RR^{n+2g},\CC\right)
\eea
with  multiplication defined by
\bee
(\xi \tens F)\cdot (\eta \tens K)=\xi\cdot _{U \eta} \tens FK+ i\hbar \eta \tens F\{\xi \tens 1,1\tens K\},
\eee
where $\xi , \eta \in   \bigoplus_{k=1}^{n+2g}\mathfrak{sl}(2,\CC)_\RR   ,$ $ F, K \in \calC^\infty\left(SL(2,\CC)_\RR^{n+2g},\CC\right)$ and $\cdot_U$ denotes the multiplication in 
$U\left( \bigoplus_{k=1}^{n+2g}\mathfrak{sl}(2,\CC)_\RR   \right)$.
The bracket $\{ , \}$ is given by (\ref{floweralgebra}).
The representation theory of this algebra is best investigated in the framework of representation theory of transformation group algebras. 
%We refer the reader to \cite{} for further details  on the general construction for a Lie the group $SL(2,\CC)_\RR$.

%\subsection{Representations of the quantum flower algebra}
The irreducible representations of the flower algebra are labelled by $n$ $SL(2,\CC)_\RR$-conjugacy classes $\calC_{\mu_i}=\{ gg_{\mu_i} g^{-1}| g\in (SL(2,\CC)_\RR\}$, $i=1,...,n,$ and the irreducible unitary Hilbert space representation $\Pi_{s_i}: N_{\mu_i}\rightarrow V_{s_i}$ of stabilisers $N_{\mu_i}=\{g\in SL(2,\CC)_\RR | gg_{\mu_i} g^{-1}=g_{\mu_i}\}$ of chosen elements $g_{\mu_1}, ..., g_{\mu_n}$ in the conjugacy classes. Consider the space 
\bea L^2_{\mu_1s_1...\mu_ns_n}&=&\{\psi: SL(2,\CC)_\RR^{n+2g}\rightarrow V_{s_1}\tens ...\tens V_{s_n}|\psi(v_1h_1,...,v_{M_1}h_n,u_{A_1}, ..., u_{B_g})\nonumber \\
&=&(\Pi_{s_1}(h_1^{-1})\tens ...\tens \Pi_{s_n}(h_n^{-1}))\psi(v_1h_1,...,v_{M_1}h_n,u_{A_1}, ..., u_{B_g})
\eea
for all $h_i\in N_{\mu_1}$ and $||\psi||^2<\infty$, with inner product 
\bee
\langle\psi,\phi\rangle=\int_{\Omega}(\psi,\phi)(v_{M_1},...,v_{M_n},u_{A_1}, ..., u_{B_g})dm_1(v_{M_1}\cdot N_1)\cdot\cdot\cdot dm_n(v_{M_n}\cdot N_n)\cdot du_{A_1}\cdot\cdot\cdot du_{B_g},
\eee
where $\Omega= SL(2,\CC)_\RR/N_{\mu_1}\times...\times SL(2,\CC)_\RR/N_{\mu_n}\times SL(2,\CC)_\RR^{2g}$  and $(\;,\;)$ is the canonical inner product. 
Then the representation spaces $V_{\mu_1s_1...\mu_ns_n}$ are obtained from $ L^2_{\mu_1s_1...\mu_ns_n}$ by removing
 the zero-norm states. The quantum flower algebra acts on a dense subspace $V_{\mu_1s_1...\mu_ns_n}$ of $C^\infty$-vectors \cite{MeusburgerSchroers2}.

\subsection{Symmetries  and the quantum double $D(SL(2,\CC)_\RR)$}

The quantum double provides a transformation group algebra associated to the puncture and handle algebras and therefore one obtains a representation of the quantum double on the representation  space of the quantum Fock-Rosly algebra for $SL(2,\CC)_\RR\rcross \Rsix$. The quantum double or Drinfeld double $D(G)$ of  a Lie group $G$ has been studied extensively, see \cite{Majidbook,CP}. 
It is a quasitriangular ribbon Hopf algebra which as a vector space can be identified as the tensor product 
$D(G)\equiv D(F(G))=F(G)\tens \CC(G)$ of the space of functions on $G$ and the group algebra $\CC(G)$.
Following conventions in \cite{KooMuller}, where the quantum is identified with continuous functions on $G\times G$ but identify $D(G)=C_0(G\times G, \CC)$ as a vector space \cite{MeusburgerSchroers2}.
The Hopf algebra structure is given as follows: The algebra (product $\bullet$ and unit $1$) is given by 
\begin{equation}
\label{alg}
 (F_1 \bullet F_2)(u,v):= \int_G F_1(u,z)F_2(z^{-1}uz,z^{-1}v)dz,\quad 1(u,v)=\delta_e(v).  
\end{equation}
The coalgebra (coproduct $\Delta$ and counit $\varepsilon$) is 
\begin{equation}
\label{coalg}
(\Delta F)(u_1,v_1;u_2,v_2) = F(u_1u_2,v_1)\delta_{v_1}(v_2), \quad \varepsilon (F)=\int _G F(e,v)dv.
\end{equation}
The antipode $S$ is 
\begin{equation}
\label{antipode}
(S F)(u,v) = F(v^{-1}u^{-1}v, v^{-1})
\end{equation}
and star structure is 
\begin{equation}
\label{star}
F^*(u,v)=\overline{F(v^{-1}uv, v^{-1})}.
\end{equation}
In terms of the singular element $f\tens \delta_g$, the relations (\ref{alg}) to (\ref{star}) take the form 
\begin{align}
&(f\tens \delta_{g_1})\bullet (f\tens \delta_{g_2}) = (f_1\cdot f_2\circ \Ad_{g_1^{-1}})\tens \delta_{{g_1}{g_2}},\\
&\Delta (f\tens \delta_g)(u_1,v_1;u_2,v_2)= f(u_1u_2)\delta_g(v_1)\delta_g(v_2),\\
&S(f\tens \delta_g)(u,v)= f(v^{-1}u^{-1}v)\delta_{g^{-1}}(v)\\
&\varepsilon(f\tens \delta_g) = f(e), \\
&(f\tens \delta_g)^*= (\bar{f}\circ \Ad_{g^{-1}})\tens \delta_{g^{-1}}
\end{align}
The Hopf algebra $D(G)$ is quasitriangular with universal $R$-matrix, 
\begin{equation}
\label{Rmatrix}
R(u_1,v_1;u_2,v_2)=\delta_e(v_1)\delta_e(u_1v_2^{-1}).
\end{equation}
Its central ribbon element $c$ is 
\begin{equation}
\label{ribbon}
c(u,v)=\delta_v(u)
\end{equation}
and satisfy the ribbon relation $\Delta c=(R_{21}\bullet R)\bullet (c\tens c)$ where $R_{21}(u_1,v_1;u_2;v_2):=R(u_2,v_2;u_1,v_1)$

The irreducible representations of $D(G)$ \cite{KooMuller} are given in are labelled by the $G$-conjugacy classes $\calC_\mu = \{v \cdot g_\mu\cdot  v^{-1}|   v \in G\}$ and irreducible unitary representations $\Pi_s$ of associated stabiliser $N_{\mu_i}=\{g\in G | gg_{\mu_i} g^{-1}=g_{\mu_i}\}$ on Hilbert spaces $V_s.$ The representation spaces are
\bea V_{\mu s}&=&\{\psi: G\rightarrow V_s |\psi(vn)=\Pi_s(n^{-1}\psi(v), \quad \forall n\in N_\mu, \forall v \in G,\nonumber \\
&&\text{and}\quad ||\psi||^2:= \int_{G/N_\mu}||\psi(z)||^2_{V_s} dm(zN_\mu)<\infty\}/\sim
\eea
where $\sim$ denotes division by zero-norm states and $dm$ is an invariant measure on
$G/N_\mu$.
The singular elements have the simple action given by
\bee
\Pi_{\mu s}(\delta_g\tens f)\psi(v)=f(vg_\mu v^{-1})\psi(g^{-1}v).
\eee
%\todo{The particular case of $G=SL(2,\CC)_\RR$}

\section{Concluding Remanrks   } 

In this paper we have studied the  (2+1)-dimensional  analog self-dual  gravity  which is obtained via dimension reduction of the  (3+1)-dimensional Holst action. This symmetry reduction consists in imposing invariance along a given spatial direction, which reduces the original four-dimensional Holst action to an action for three-dimensional gravity with a Barbero-Immirzi parameter.  In particular, we constructed a Chern-Simons  theory for this model based on the gauge group $SL(2,\CC)_\RR\rcross \Rsix$. The idea for this was to view 
 the  $3d$ complex self-dual $\mathfrak{sl}(2,\RR)_\CC$  variables to 
 $6d$ real variables  which combines into a $12d$ Cartan connecttion.
 \\
 
 We then apply the  combinatorial quantisation program to this  Chern-Simons theory. 
 Here, the description of the phase space of Chern-Simons gauge theory with the semidirect product gauge groups $H =SL(2,\CC)_\RR\rcross \Rsix$ on a punctured surface is 
  given by the Fock-Rosly construction, which is a crucial ingredient in 
 The Poisson structure for the moduli space of flat % $SL(2,\CC)_\RR\rcross \Rsix$-
 connections is given in terms of the  classical $r$-matrix for the quantum double $D(SL(2,\CC)_\RR)$ viewed as the double of a double $ D(SL(2,\RR)\dcross AN(2))$. This  quantum double 
 provides a feature for quantum symmetries of the quantum theory for this model.
 One can argue that, in the context of the combinatorial quantisation program, the implementation of reality conditions amount to extending the symmetries to higher dimension. In this case, 12 dimensional symmetry.
 \\

It is important to note that the (2+1)-dimensional  analog self-dual  gravity studied here provides a testing ground for understanding the implementation of the so called reality condtions arising when working with the complex connection in loop quantum gravity, i.e the three dimensional analog of the self dual Ashtekar connection. 
An investigation of this and how the results compare with the combinatorial formalism described in the work is deferred for future work.
\\

In the standard application of the combinatorial framework to the Chern-Simons theory for (2+1)-dimensional gravity, what emerges is that  quantisation deforms the  classical phase space  geometry into a non-commutative geometry in which the  model space-times are replaced by non-commutative spaces and the local  isometry groups (gauge groups for the Chern-Simons theory) by quantum groups. In the current construction, the role of the quantum groups, i.e. the quantum double $D(SL(2,\CC)_\RR)$ seem to be mathematical objective for construction the Hilbert Space for the quantum theory. 
However, we expect the quantum groups  to act on some deformation of the space-time. 
The general Chern-Simons theory proposed here in \eqref{SCgamma} for the inclusion of the Barbero-Immirzi parameter in the Euclidean setting require further investigation.

\section*{Acknowledgments}
I  wish to thank the Karim Noui, Jibril~Ben Achour and  Bernd~J.~Schroers for illuminating discussion.

\appendix
\section{A decomposition of the 3d self-dual action}
\label{analysis with gamma complex}
We consider the analysis for the complex-valued action 
\begin{equation}
\label{Caction}
S_{}[\chi,e,\omega]= \int_{M^3} \textbf{ d}^3x\epsilon^{\mu\nu\rho}\left(\frac{ 1}{ 2 }\epsilon_{IJKL} \chi^I e^J_{\mu} F^{KL}_{\nu\rho} \osign  i\chi^Ie^J_\mu F_{\nu\rho IJ}\right)
\end{equation}
obtained from \eqref{HRaction} by setting $\gamma = i$.
 The 2+1 space-time decomposition of (\ref{Caction}) gives 
\begin{align}
\label{Cspace-time}
S[x,e,\omega]= & \int_{M^3} \textbf{ d}^3x\epsilon^{\mu\nu\rho}\left(\frac{ 1}{ 2 }\epsilon_{IJKL} \chi^I e^J_{\mu} F^{KL}_{\nu\rho} \osign i \chi^Ie^J_\mu F_{\nu\rho IJ}\right) \nonumber \\ 
= & \int_{M^3} \textbf{ d}^3x\left(\epsilon^{0ab} (\frac{ 1}{ 2 }\epsilon_{IJKL} \chi^I e^J_{0} F^{KL}_{ab} \osign i \chi^Ie^J_0 F_{ab IJ})+\epsilon^{ab0}(\epsilon_{IJKL} \chi^I e^J_{a} F^{KL}_{b0} \osign 2i \chi^Ie^J_a F_{b0 IJ})\right) \nonumber \\
= &\frac{1}{ 2} \int_{M^3} \textbf{ d}^3x\left(2\epsilon^{ab}\chi^Ie^J_a (\epsilon_{IJKL}  F^{KL}_{b0} \osign 2i  F_{b0 IJ})+\epsilon^{ab} \chi^I e^J_{0}(\epsilon_{IJKL} F^{KL}_{ab} \osign 2i  F_{ab IJ})\right)\nonumber \\ 
=&\frac{1}{ 2} \int_{M^3} \textbf{ d}^3x\left(L_C+L_S \right).
\end{align}
In the 3+1 internal gauge group decomposition, we have 
\begin{align}
L_C=&2\epsilon^{ab}(\chi^0e_a^j-\chi^je_a^0)(\epsilon_{jkl} F^{kl}_{b0} \osign 2i  F_{b0\, 0j})+4\epsilon^{ab}\chi^j e^k_a (\epsilon_{jkl} F^{0l}_{b0} \osign i  F_{b0\, jk})\nonumber \\
=&2\epsilon^{ab}E_a^{[0j]} (\epsilon_{jkl} F^{kl}_{b0} \osign 2i  F_{b0 \,0j})+4\epsilon^{ab}E_b^{jk} (\epsilon_{jkl} F^{0l}_{b0} \osign i  F_{b0 \,jk})
\end{align}
and
\begin{align}
L_S=&\epsilon^{ab}(\chi^0e_0^j-\chi^je_0^0)(\epsilon_{jkl} F^{kl}_{ab} \osign 2i  F_{ab\, 0j})+2\epsilon^{ab}\chi^j e^k_0 (\epsilon_{jkl} F^{0l}_{ab} \osign i  F_{ab\, jk})\nonumber \\
=&\epsilon^{ab} E_0^{[0j]}  (\epsilon_{jkl} F^{kl}_{ab} \osign 2i  F_{ab\, 0j})+2\epsilon^{ab}E_0^{jk} (\epsilon_{jkl} F^{0l}_{ab} \osign i  F_{ab \,jk}),
\end{align}
where
\begin{equation}
E_\mu^{[0j]} = (\chi^0e_\mu^j-\chi^je_\mu^0), \quad E_\mu^{kj} =\chi^j e^k_\mu.
\end{equation}

Now the curvature $F^{IJ}$ of the spin connection $\omega^{IJ}$ given in \eqref{curvature} can be decomposed into
\begin{align}
F^{0l}_{\mu\nu}
=&\partial_\mu \omega^{0l}_{\nu}-\partial_\nu\omega^{0l}_\mu +(\omega_\nu\times\omega_\mu^{(0)})^l-(\omega_\mu\times\omega_\nu^{(0)})^l
\end{align}
and 
\begin{align}
F^{jk}_{\mu\nu}
=&\epsilon^{jk}\,_l\left(\partial_\mu \omega^{l}_{\nu}-\partial_\nu\omega^{l}_\mu -(\omega_\mu\times\omega_\nu)^l+(\omega^{(0)}_\mu\times\omega^{(0)}_\nu )^l\right)
\end{align}
so that 
\begin{align}
F^{k}_{\mu\nu}=&\frac 1 2 \epsilon^{k}_{\;ij}F^{ij}_{\mu\nu} = \partial_\mu \omega^{k}_{\nu}-\partial_\nu\omega^{k}_\mu -(\omega_\mu\times\omega_\nu)^k+(\omega^{(0)}_\mu\times\omega^{(0)}_\nu )^k
\end{align}
where we have used $ \omega_\mu^i \equiv \epsilon^i_{\,jk}\omega^{jk}_\mu .$
Therefore we have that
\begin{align}\label{curvatures}
\nonumber
\epsilon_{jkl} F^{kl}_{b0} \osign 2i  F_{b0 0j}=&\epsilon_{jkl} F^{kl}_{b0} \sign2i  F_{b0}^{0j}=2\left[\partial_b A^{j}_{0}-\partial_0A^{j}_b +(A_0\times A_b)^j\right],\\\nonumber
\epsilon_{jkl} F^{0j}_{b0} \osign i  F_{b0 kl}= & \osign i\epsilon_{jkl}\left[\partial_b A^{j}_{0}-\partial_0A^{j}_b +(A_0\times A_b)^j\right],
\\\nonumber
\epsilon_{jkl} F^{kl}_{ab} \osign 2i  F_{ab 0j}=&\epsilon_{jkl} F^{kl}_{ab} \osign 2i  F_{ab}^{0j}=2\left[\partial_a A^{j}_{b}-\partial_bA^{j}_a -(A_a\times A_b)^j\right] \\ %\nonumber
\epsilon_{jkl} F^{0j}_{ab} \osign i  F_{ab kl}=&\epsilon_{jkl} F^{0j}_{ab} \osign i  F_{ab}^{ kl}=  \osign i\epsilon_{jkl}\left[\partial_a A^{j}_{b}-\partial_bA^{j}_a -(A_a\times A_b)^j\right],
\end{align}
where
\begin{equation}
A_\mu^{j} = \omega_\mu^j\sign i\omega^{(0)j}_\mu=\omega_\mu^j\sign i\Gamma^{j}_\mu.
\end{equation}
Using the relations (\ref{curvatures}) in \eqref{Cspace-time}, we obatain
\begin{align}
L_C=&4E_a^{[0j]}\epsilon^{ab} \left[\partial_b A^{j}_{0}-\partial_0A^{j}_b +(A_0\times A_b)^j\right] \sign 4i\epsilon_{jkl} E_a^{kl}\epsilon^{ab} \left[\partial_b A^{j}_{0}-\partial_0A^{j}_b +(A_0\times A_b)^j\right]\nonumber\\
=&4 \epsilon^{ab}\left((E_a^{[0j]} \sign  i \epsilon_{jkl}E_a^{kl}) \partial_0A^{j}_b-( E_a^{[0j]} \sign i \epsilon_{jkl} E_a^{kl} )\partial_b A^{j}_{0}-(E_a^{[0j]} \sign i \epsilon_{jkl}E_a^{kl})(A_0\times A_b)^j\right)\nonumber\\
=&4\epsilon^{ab}\left(E_a^j \,\partial_0A^{j}_b-E_a^j\partial_b A^{j}_{0}-E_a^j(A_0\times A_b)^j\right)
\end{align}
and
\begin{align}
L_S=&2E_0^{[0j]}\epsilon^{ab}\left[\partial_a A^{j}_{b}-\partial_bA^{j}_a -(A_a\times A_b)^j\right] \sign 2 i\epsilon_{jkl}E_0^{kl}\epsilon^{ab}\left[\partial_a A^{j}_{b}-\partial_bA^{j}_a -(A_a\times A_b)^j\right]  \nonumber\\
=&2\left(E_{0}^{[0j]} \sign  i  \epsilon_{jkl}E_0^{kl}\right)\epsilon^{ab}\left[\partial_a A^{j}_{b}-\partial_bA^{j}_a -(A_a\times A_b)^j\right] 
\nonumber\\
=&2E_0^j\epsilon^{ab}\left(\partial_a A^{j}_{b}-\partial_bA^{j}_a -(A_a\times A_b)^j\right),
\end{align}
where
\begin{equation}
 E_\mu^j=E_{\mu}^{[0j]} \sign i  \epsilon^{j}_{kl}E_{\mu}^{\,{kl}}%=\calB_{\mu}^j \sign  i \epsilon^{jkl}B_{\mu\,{kl}}.
\end{equation}
Thus, the action (\ref{Caction}) becomes
\begin{align}
\nonumber \label{decCaction}
S[E, A]= & \int_{M^3} \textbf{d}^2x\textbf{d}x^0 \epsilon^{ab}   \left( 2\left(E_a \cdot\partial_0A_b+A_0\cdot(\partial_bE_a-A_b\times E_a)\right) 
+E_0\cdot \left(\partial_a A_{b}-\partial_bA_a -A_a\times A_b\right)\right),\\\nonumber
=&\int_{M^3}\textbf{ d}^3x\,\epsilon^{\mu\nu\rho}  E_\mu F_{\nu\rho} 
 = \int_{M^3}  E \wedge F[A],
\end{align}
together with the effective  self-dual variables
\begin{equation} \label{SDvariable}
 E_\mu^j=(\chi^0e_\mu^j-\chi^je_\mu^0) + i  \epsilon^{j}_{kl}\,\chi^k e^l_\mu,
 %E_{\mu}^{[0j]} \sign i  \epsilon^{j}_{kl}E_{\mu}^{\,{kl}},
  \quad A_\mu^{j} = \omega_\mu^j\sign i\omega^{(0)j}_\mu.
\end{equation}
%The curvature of the self-dual Ashteker  connection $A_a$ introduced here is 
%\bee
%F(A)=\partial_a A_{b}-\partial_bA_a -A_a\times A_b .
%\eee

\newpage

\bibliographystyle{plain}
\bibliography{References}

\end{document}